\documentclass{aa}  
\usepackage{graphicx}

\usepackage{natbib,twoopt}
\usepackage[breaklinks=true]{hyperref} 
\bibpunct{(}{)}{;}{a}{}{,}             
\makeatletter
  \newcommandtwoopt{\citeads}[3][][]{\href{http://adsabs.harvard.edu/abs/#3}%
    {\def\hyper@linkstart##1##2{}%
     \let\hyper@linkend\@empty\citealp[#1][#2]{#3}}}
  \newcommandtwoopt{\citepads}[3][][]{\href{http://adsabs.harvard.edu/abs/#3}%
    {\def\hyper@linkstart##1##2{}%
     \let\hyper@linkend\@empty\citep[#1][#2]{#3}}}
  \newcommandtwoopt{\citetads}[3][][]{\href{http://adsabs.harvard.edu/abs/#3}%
    {\def\hyper@linkstart##1##2{}%
     \let\hyper@linkend\@empty\citet[#1][#2]{#3}}}
  \newcommandtwoopt{\citeyearads}[3][][]%
    {\href{http://adsabs.harvard.edu/abs/#3}
    {\def\hyper@linkstart##1##2{}%
     \let\hyper@linkend\@empty\citeyear[#1][#2]{#3}}}
\makeatother

\newcommand{\ind}[1]{_{\rm #1}}

\begin{document} 

\title{Cold and warm electrons at comet 67P}

   \author{A. I. Eriksson\inst{1}
  \and
  I. A. D. Engelhardt\inst{1,2}
    \and
  M. Andr\'{e}\inst{1}
      \and 
  R. Bostr\"{o}m\inst{1}
    \and 
  N. J. T. Edberg\inst{1}
   \and \\
  F. L. Johansson\inst{1,2}
  \and 
  E. Odelstad\inst{1,2}
  \and 
  E. Vigren\inst{1}
  \and 
  J.-E. Wahlund\inst{1}
  \and
  P. Henri\inst{3}  
  \and
  J.-P. Lebreton\inst{3}
  \and\\
  W. J. Miloch\inst{4}
  \and
  J. J. P. Paulsson\inst{4}
  \and
  C. Simon Wedlund\inst{4}
  \and
  L. Yang\inst{4}
  \and
  T. Karlsson\inst{5}
  \and
  R. Jarvinen\inst{6}
  \and\\
  T. Broiles\inst{7}
    \and
  K. Mandt\inst{7,8}      
  \and
  C. M. Carr\inst{9}  
  \and
    M. Galand\inst{9}
  \and
  H. Nilsson\inst{10}
  \and
  C. Norberg\inst{10}
  }

\institute{
Swedish Institute of Space Physics, Uppsala, Sweden
\and Department of Physics and Astronomy, Uppsala University, Sweden
\and LPC2E, Orl\'{e}ans, France
\and Department of Physics, University of Oslo, Norway
\and Alfv\'{e}n Laboratory, Royal Institute of Technology, Stockholm, Sweden
\and Finnish Meteorological Institute, Helsinki, Finland
\and Southwest Research Institute, San Antonio, TX, USA
\and Department of Physics and Astronomy, University of Texas at San Antonio, San Antonio, TX, USA
\and Department of Physics, Imperial College London, UK
\and Swedish Institute of Space Physics, Kiruna, Sweden
}


 
\abstract
   {Strong electron cooling on the neutral gas in cometary comae has been predicted for a long time, but actual measurements of low electron temperature are scarce.}
   {Our aim is to demonstrate the existence of cold electrons in the inner coma of comet 67P and show filamentation of this plasma.}
   {In situ measurements of plasma density, electron temperature and spacecraft potential were carried out by the Rosetta Langmuir probe instrument, LAP. We also performed analytical modelling of the expanding two-temperature electron gas.}
   {LAP data acquired within a few hundred km from the nucleus are dominated by a warm component with electron temperature typically 5--10~eV at all heliocentric distances covered (1.25 to 3.83~AU). A cold component, with temperature no higher than about 0.1~eV, appears in the data as short (few to few tens of seconds) pulses of high probe current, indicating local enhancement of plasma density as well as a decrease in electron temperature. These pulses first appeared around 3~AU and were seen for longer periods close to perihelion. The general pattern of pulse appearance follows that of neutral gas and plasma density. We have not identified any periods with only cold electrons present. The electron flux to Rosetta was always dominated by higher energies, driving the spacecraft potential to order $-10$~V.}
   {The warm (5--10~eV) electron population observed throughout the mission is interpreted as electrons retaining the energy they obtained when released in the ionisation process. The sometimes observed cold populations with electron temperatures below 0.1~eV verify collisional cooling in the coma. The cold electrons were only observed together with the warm population. The general appearance of the cold population appears to be consistent with a Haser-like model, implicitly supporting also the coupling of ions to the neutral gas. The expanding cold plasma is unstable, forming filaments that we observe as pulses.}


\maketitle
%


\section{Introduction}

When the Rosetta spacecraft arrived at comet 67P/Churyumov-Gerasimenko at 3.6~AU heliocentric distance, the plasma environment within some hundred kilometer distance was already dominated by cometary matter \citep{Yang2016a}. The plasma density and spacecraft potential were modulated by the nucleus spin period in a pattern that persisted for all observations in the northern hemisphere of the comet at least during northern summer and followed the density variations of the neutral gas \citep{Edberg2015a,Odelstad2015a}. Within a few 100~km distance of the nucleus, the main determinant for the cometary plasma bulk properties from then on was the nucleus outgassing, though solar wind variations certainly had an impact \citep{Edberg2016a,Edberg2016b}. Comet plasma environments and their interactions with the solar wind are often discussed as being typical for an "active comet" or for a "bare nucleus" interaction supposedly characteristic of asteroids, where the few cometary ions behave like test particles in the solar wind \citep{Coates1997a,Coates2009a}. 67P is not a very active comet \citep{Almeida2009a,Bieler2015a,Gulkis2015a,Snodgrass2016b}, but even at 3.6~AU upon Rosetta's arrival on the inbound leg of the comet orbit or at 3.83~AU on the outbound leg at the end of the mission, 67P did not behave as a bare asteroid. From a plasma point of view, the comet was active throughout the Rosetta mission. 

The comet ionosphere is formed from the atmosphere mainly by photoionisation, charge exchange with solar wind ions, and, at least at times, also by impact ionisation by high-energy electrons \citep{Cravens1987b,Galand2016a}. The typical energy of recently released photoelectrons is expected to be 12--15~eV \citep{Haberli1996a,Galand2016a}. As the ionospheric plasma at a comet is not stationary but expanding, adiabatic cooling will play some role, but in a sufficiently dense neutral gas, the main cooling agent for electrons will be collisions with neutrals. The neutral gas is expected to have a temperature $T\ind{n}$ of a few hundred K or even less \citep[e.g.\,][]{Tenishev2008a}. For highly active comets, like 1P/Halley at perihelion, collisional cooling of the electrons is expected to be so efficient as to keep $T\ind{e}$ close to $T\ind{n}$ out to several thousand kilometers \citep{Gan1990a}. 

While models more or less unanimously agree that the cometary electron gas experiences cooling, actual reports of $T\ind{e}$ below the approximately 10~eV expected for solar wind, as well as recent photoionisation products, are sparse. Giotto could not reliably access electrons below about 10~eV, so only indirect evidence of cold electrons is available from its encounter with 1P/Halley. A plasma density change near 15,000~km was interpreted as indirect evidence of an electron collisionopause, a boundary between collisional (efficient electron cooling) and collisionless regimes for the electrons \citep{Ip1986a,Gan1990a,Haberli1996a}, but this provides little direct information on $T\ind{e}$. The Vega spacecraft, also visiting 1P in 1986, included a Langmuir probe instrument, from which \citet{Grard1989a} infered an electron temperature of around 0.5~eV from a distance of around 900,000~km in to 29,000~km (Vega~1) and 66,000~km (Vega~2). Finally, \citet{Meyer-Vernet1986a} used thermal noise measurements by the International Cometary Explorer (ICE) during its crossing of the tail of comet 21P/Giacobini-Zinner to demonstrate how $T\ind{e}$ decreased from around 10~eV in the outer reaches of the tail to 1~eV in the central tail at a closest approach distance to the nucleus of 7,800~km. Prior to Rosetta, there were no $T\ind{e}$ measurements in the innermost coma.

Rosetta followed comet 67P/Churyumov-Gerasimenko in its orbit for more than two years, staying within a few hundred kilometres  except for two excursions, and even for several months within a few tens of kilometres. In order of magnitude, these distances are about the same in nucleus radii as in kilometers, as $r\ind{67P} = 1.65$~km if defined as the radius of a sphere of volume $18.8 \pm 0.3$~km$^3$ found for the 67P nucleus by \citet{Jorda2016a}. Heliocentric distance varied from 3.6~AU at arrival to 1.25~AU at perihelion, and out again to 3.83~AU at the end of the mission. The plasma instruments onboard were operational almost all the time, creating a database of cometary plasma measurements vastly larger and over a wider activity range than any previous mission. This clearly gives good opportunities for detecting cold electrons and for following their evolution during varying stages of comet activity.

As 67P has about 1\% of the production rate of 1P \citep{Almeida2009a}, it is not obvious how efficient electron cooling should be, particularly during the early and late mission stages far from perihelion. \citet{Mandt2016a} used measured daily averages of neutral gas density inside 1.9~AU to show that while Rosetta has spent most of its time in a region where the ions collisionally couple to the neutral gas, the local electron-neutral collision rate at Rosetta was for most of the time insufficient for effective cooling of electrons. A similar conclusion was reached by \citet{Galand2016a} for 3~AU. This is consistent with the negative spacecraft potential \citep{Odelstad2015a}, which requires a substantial flux of electrons in an energy range comparable to or above the spacecraft potential. However, Figure~5 of \citet{Mandt2016a} indicates that at least for heliocentric distances inside 1.9~AU  (meaning from April~2015), a region where collisional cooling of electrons is important should exist close to the nucleus. 

If transported with the neutral flow, we would expect these cold electrons to reach Rosetta also when the spacecraft is outside this region, unless heating and recombination processes act sufficiently fast to destroy such a cold population. Evidence for the plasma flow following the neutrals comes from the observed bulk density of electrons. \citet{Edberg2015a} noted that during a flyby in February~2015 (at 2.3~AU), the plasma density approximately followed the $1/r$ dependence on cometocentric distance $r$ predicted if the plasma and neutral gas expansion speeds are equal and constant \citep{Haser1957a}. Furthermore, \citet{Odelstad2015a}, \citet{Vigren2015b}, and \citet{Galand2016a} have shown that from early in the mission, diurnal as well as hemispherical plasma density variations follow the neutral gas, with local ionisation provided by solar EUV and sometimes by high-energy electrons. In addition, ion observations \citep{Nilsson2015a,Nilsson2015b,Goldstein2015a,Broiles2015a} show two major populations of cometary ions: Partially picked up cometary ions accelerated to a fraction (hundreds of eV) of their final energy when Rosetta observes them, and low energy ions entering the ion detectors at energies not much above that the acceleration towards the detectors by the negative spacecraft potential would give them (up to a few tens of eV, compared to the 0.1~eV kinetic energy of a water ion drifting with the neutral gas at 1~km/s). The latter population should be the bulk plasma ions, kept at low energy by the coupling to the expanding neutral gas. We therefore conclude that in the intervals covered by these studies, at least a substantial fraction of the ions are indeed collisionally bound to the neutral gas, agreeing with the conclusion by \citet{Mandt2016a} and \citet{Galand2016a} that Rosetta is mostly in a region where ion-neutral coupling is substantial. 

The coupling of the plasma to the neutral gas breaks down at scales below the ion collision length. In contrast to the rather smooth neutral density time series observed by Rosetta \citep{Hassig2015a,Bieler2015a,Hansen2016a}, the plasma density therefore can show strong variations. For example, the $\sim 1/r$ plasma density profile shown by \citet{Edberg2015a} displays significant relative variations around this mean profile, as large as an order of magnitude or even more. The comparisons of plasma to neutral gas density by \citet{Odelstad2015a}, \citet{Vigren2015b}, and \citet{Galand2016a} also show much stronger density fluctuations in the plasma than in the neutral gas. For the plasma, expanding in a strong density gradient and with substantial shear flows, one expects low-frequency instabilities and structure formation \citep{Ershkovich1988a,Thomas1995a,Rubin2012a,Koenders2015a}. Furthermore, waves at frequencies capable of electron heating may also be present; for example, lower hybrid waves. \citet{Broiles2016b} suggest wave heating by lower hybrid waves may play a role in the energisation of electrons seen on Rosetta at above 10~eV, in addition to the particle kinetic effects modelled by \citet{Madanian2016a}. Such waves have indeed been observed at 67P \citep{Karlsson2017a}, though their role in energy transport between various particle populations remains to be investigated in detail.

From the above, we may conclude that while Rosetta is expected to be mostly outside the region of efficient electron cooling close to the nucleus,
there could be a significant population of cold electrons reaching Rosetta's location, in addition to the warmer population driving the spacecraft potential negative. In this report we use data from the Rosetta Langmuir probe instrument LAP to look for signatures of these cool electrons. Section~\ref{sec:instr} presents the instrument, its measurements, and data interpretation issues in various plasma density regimes. In Section~\ref{sec:cold} we discuss cold electron observations, and present examples and statistics of how cold plasma often appears in pulses. Physical interpretation and model comparisons of these observations are then the topics of Section~\ref{sec:disc}, before a concluding discussion in Section~\ref{sec:concl}.


\section{Instrumentation and data}\label{sec:instr}

The Rosetta Plasma Consortium (RPC) \citep{Carr2007a} comprises a set of instruments together observing the fundamental parameters of the cometary plasma. This paper is based on data from the dual Langmuir probe instrument (LAP), whose main objective is to gain measurements of the bulk parameters of the cometary plasma, particularly its density. For context, we also refer to data from the COPS neutral gas pressure and density sensor of the ROSINA instrument \citep{Balsiger2007a,Bieler2015a}.

\begin{figure}
\centering
\includegraphics[width=\columnwidth]{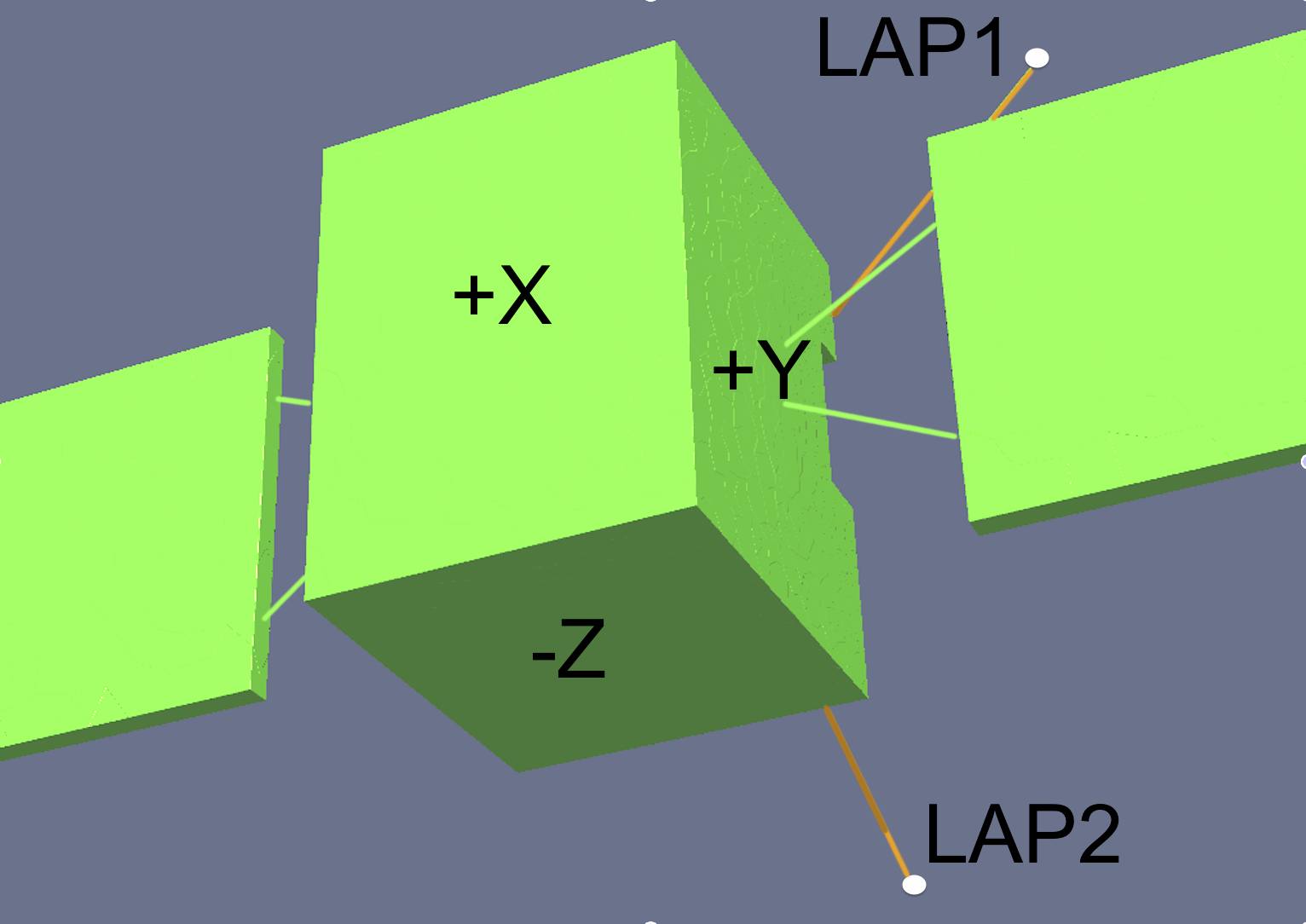}
\caption{Schematic showing the mounting of the LAP sensors. The LAP1 boom is in the YZ plane while the LAP2 boom is close to the XZ plane. Most of the time +Z pointed to the comet nucleus and +X was roughly in the solar direction. The solar panels and hence the Y axis were almost always perpendicular to the Sun. Not included is the high-gain antenna, mounted on the -Z side.}
\label{fig:rpc}
\end{figure}

We first summarize some main properties of LAP in Section~\ref{sec:lap}, turning to models for the interpretation of current-voltage characteristics in various environments in Section~\ref{sec:interp}. For the Rosetta mission, the plasma environments encountered can be grouped into three classes of plasma regime as below, roughly corresponding to plasma density ranges as indicated in Sections~\ref{sec:tenuous} -- \ref{sec:dense}.

Unless otherwise indicated, we assume in this Section that all particle species are described by single Maxwellian distribution functions, that the charge density on dust grains is negligible compared to that of free ions and electrons, and that all ions are singly charged. Effects of violation of these assumptions will be raised as we present actual data in Section~\ref{sec:cold}.

\subsection{The Langmuir probe instrument (LAP)}\label{sec:lap}

The main task of LAP is to investigate the properties and dynamics of the cometary plasma environment through some of its principal bulk properties, the number density $n$ and electron temperature $T\ind{e}$, as well as the spacecraft potential $V\ind{s}$. We use the term temperature rather freely, assuming Boltzmann-like energy distributions in some energy range rather than perfect equilibrium distributions. Depending on operational mode and plasma environment, LAP can in some conditions also provide ion flow velocity or temperature, an effective ion mass, an integrated solar EUV flux measure from the observed photoemission, and an electric field component estimate for low-frequency oscillations, and plasma waves up to 8~kHz. 

To these ends, LAP uses two spherical sensors of 50~mm diameter mounted at the tips of two booms asymmetrically protruding from the spacecraft body (Figure~\ref{fig:rpc}).  The basic measured property is the current flowing from the probe to space due to collection of various particle species in the plasma and emission of photoelectrons and secondary electrons when a bias voltage is applied to the probe. An alternative mode, where the probes are fed with a bias current and their voltages are measured, is not used in this paper. Typical operations of the instrument combine probe bias voltage sweeps for obtaining the Langmuir probe characteristic (I-V curve) at intervals of a few minutes, with continuous sampling (at around 1~Hz or up to 58~Hz sampling frequency, depending on available telemetry rate) of probe current at constant positive or negative bias voltage to cover the plasma dynamics between sweeps. Full instrument descriptions are provided by \citet{Eriksson2007d} and \citet{Eriksson2008a}.

The two booms hold the LAP1 and LAP2 probes at distances from the spacecraft (boom hinge) of 2.2 and 1.6~m, respectively, while the spacecraft body itself roughly measures 2x2x3~m with solar arrays extending to a total wing span of 32~m. Due to the small distance between the spacecraft and the probes, the data can be expected to show disturbances due to spacecraft-plasma interaction, as investigated by \citet{Sjogren2012a} and \citet{Johansson2016a} and discussed below. Nevertheless, LAP allows access to plasma parameters over a very wide parameter range using techniques adapted to the environment. 

The main tool for accessing the plasma parameters by a Langmuir probe is the bias voltage sweep, in which we vary the bias potential $V\ind{b}$ between spacecraft ground and the probe and measure the resulting current $I\ind{p}$, defined as positive when flowing from the probe to the plasma. Values of plasma parameters then follow from fitting to the probe current models described in Section~\ref{sec:interp}. Sweep examples and parameter fits at various stages of the mission are shown in Figure~\ref{fig:sweeps}. The individual sweeps and the interpretation of the data in each case are discussed in Sections~\ref{sec:tenuous}--\ref{sec:cold}.

\subsection{Probe-sweep models}\label{sec:interp}

To extract information about the plasma parameters, the probe characteristic must be interpreted by physical models of the collected current from various particle populations. For the examples in this study, we consider the total probe current $I\ind{p}$ as the sum of three such fluxes: the currents carried by collection of plasma ions and electrons, $I\ind{i}$ and $I\ind{e}$, and the current due to photoemission of the probe surface itself, $I\ind{f}$. The current $I\ind{s}$ due to secondary electron emission caused by plasma electrons impacting on the probe is thus not modelled here, though it may be important at times \citep{Garnier2012a,Wang2015a}. When the spacecraft is negatively charged, as it mostly is \cite{Odelstad2015a}, the photoelectrons it produces are accelerated away from it, so their number density is much below that of the natural plasma. Only in the most tenuous plasmas encountered do we need to take their contribution to the probe current into account (Section~\ref{sec:tenuous}).

For LAP on Rosetta, the relevant framework for describing these currents is provided by the model for orbital motion limited (OML) current collection introduced by \citet{Mott-Smith1926a}, refined and much used in many variants ever since. The basic requirement is that the probe radius $a$ must be small compared to the Debye length, $\lambda_D$, though as shown by \citet[][Table 5c]{Laframboise1966a}, the OML expressions work well even for the case $a = \lambda_D$ (overestimating the collected electron current by less than 10\% for probe potentials up to ten times $KT\ind{e}/e$). Issues with this criterion may need consideration regarding electron collection in dense cold plasmas at perihelion but not otherwise. As the electron gyroradius is always much larger than the probe size, magnetic field effects on probe current collection need not be considered \citep{Laframboise1993a}. Effects of magnetic connection to spacecraft surfaces \citep{Hilgers1992a}, however, may possibly set in at the very strongest magnetic fields and lowest temperatures observed. 

\begin{figure}
\centering
\includegraphics[width=\columnwidth]{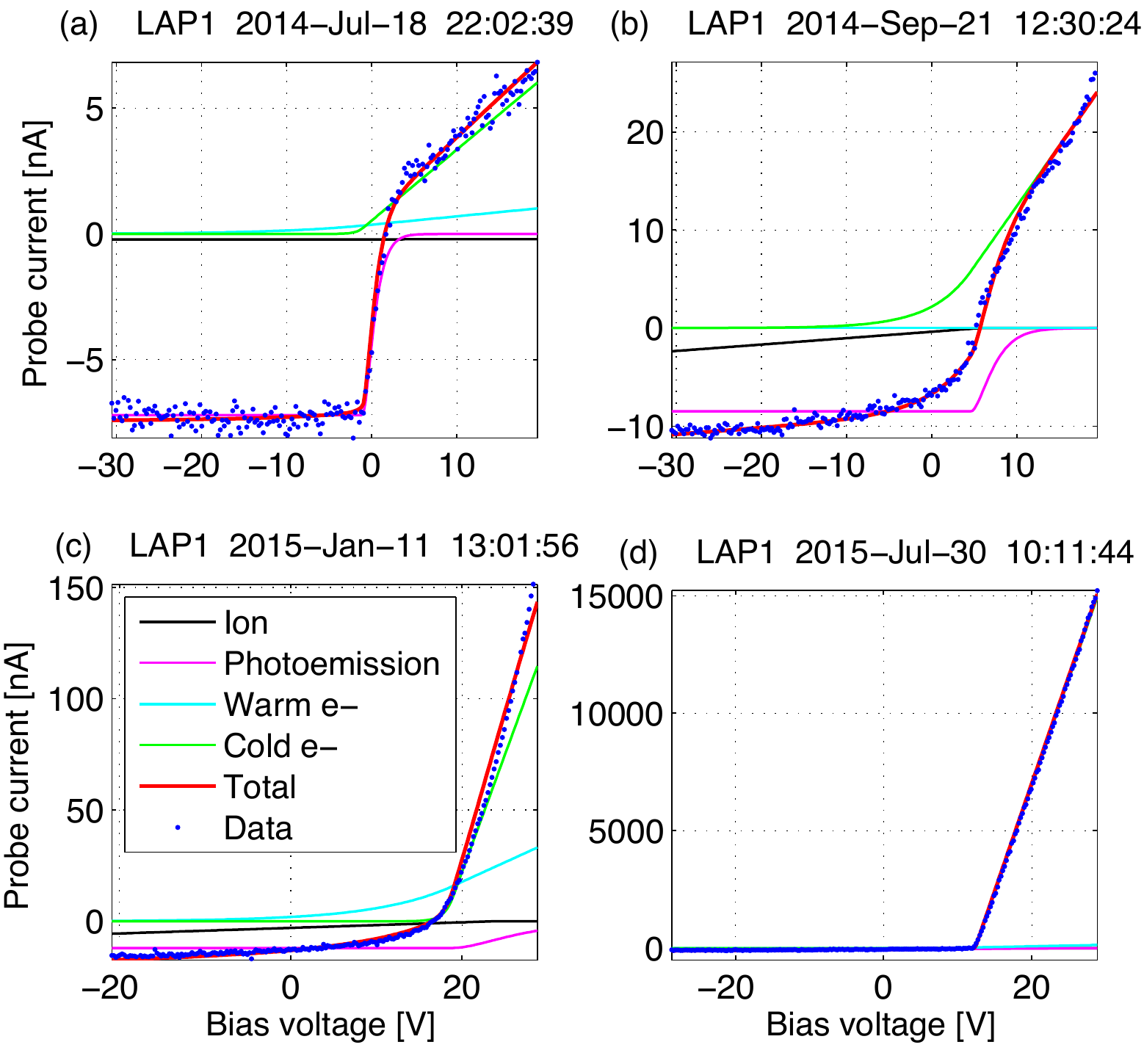}
\caption{Examples of LAP probe characteristics from various mission phases, with parameters as in Table~\ref{tab:fit} fitted to 
OML current models. Distances in AU are heliocentric and those in km are cometocentric. Data are blue dots and fitted currents are given by colours as in the box in panel (c). Panels: (a) Solar wind prior to Rosetta arrival at comet (3.7~AU, 7,000~km). (b) Warm electrons at early activity (3.33~AU, 35~km). (c) Cold electrons mix in as activity increases (2.57~AU, 28~km). (d) Dense cold plasma near perihelion (1.25~AU, 180~km).}
\label{fig:sweeps}
\end{figure}

\begin{table}
\centering
\caption{Parameters for the model fits in Figure~\ref{fig:sweeps}. Heliocentric distance $R\ind{H}$, cometocentric distance $r$ and COPS neutral gas density are given for context. The applicability of the fitted model for each example, and hence the interpretation of the parameters, is discussed in the text. Values in parentheses are not considered to accurately represent the actual plasma. Water ions are assumed for all cases except for the first. COPS density values are taken from \citet{Bieler2015a}, \citet{Vigren2016a} and \citet{Goetz2016a}.}\label{tab:fit}
\resizebox{\columnwidth}{!}{%
\begin{tabular}{|l|cccc|c|} 
                \hline
                & \multicolumn{4}{|c|}{Fig.~\ref{fig:sweeps}} & Fig.~\ref{fig:swp-pulse}\\
                & (a) & (b) & (c) & (d)\\
                \hline
                Date [yymmdd]   & 140718 & 140921 & 150111 & 150730 & 150110\\
                $R\ind{H}$ [AU]    &   3.7      & 3.33  & 2.57  & 1.25 & 2.56\\
                $r$ [km]                &   7000        & 35            & 28              & 180 & 28\\
                \hline
                $n\ind{n}$ [$10^7$ cm$^{-3}$] & ... & $ <3$ & 6.6 & 6.6 & 5 \\
                \hline
                $-V\ind{k}$ [V]                 & 1.1  & -4.6  & -18.9 & -12.4 & (-25.8)\\
                $n\ind{ew}$ [cm$^{-3}$] & 0.5  & 12.7 & 25     & 60 & (120)\\
                $T\ind{ew}$ [eV]                & 10  & 4.7    & 9       & 6 & 9 \\
                $n\ind{ec}$ [cm$^{-3}$] & 1.4  & ...       & (50)     & (1100) & (90) \\
                $T\ind{ec}$ [eV]                & 1.2  & ...        & 1       & (0.065) & (0.02) \\
                $u$ [km/s]                      & 400 & 0.65  & (2.0)    & (2.2) & (4.5) \\
                $m\ind{i}$ [amu]           & 1.06 & 18     & (18)     & (18) & (18) \\
                $I\ind{f0}$ [nA]              & 7.2  & 8.5    & 12      & 26 & 12 \\
                \hline
\end{tabular}
}
\end{table}

Following \citet{Grard1973a}, we model the current due to photoemission from a probe at potential $U\ind{p}$ with respect to its immediate surroundings as 
\begin{align} \label{if}
I\ind{f} = 
\begin{cases}
-I\ind{f0}, & U\ind{p} < 0 \\ 
-I\ind{f0} \left(1 + \frac{eU\ind{p}}{KT\ind{f}} \right)^\mu \exp \left(-\frac{eU\ind{p}}{KT\ind{f}} \right), & U\ind{p} > 0, 
\end{cases}
\end{align}  
where $I\ind{f0} = A\ind{f} j\ind{f0}$ is the photoemission saturation current, set by the probe area $A\ind{f} = \pi a^2$ projected to the sun and the photoemission current density $j\ind{f0}$, which depends on material properties as well as on the solar UV spectrum. We use the usual convention of considering actually flowing currents as positive in the direction from the probe to the plasma, though current densities such as $j\ind{f0}$ and constants like $I\ind{f0}$ represent magnitudes and are thus always positive. The exponent $\mu$ depends on what angular distribution is assumed for the emitted photoelectrons at the surface; if this is isotropic, as we assume here, $\mu = 0$, while $\mu = 1$ for purely radial emission. The model assumes a Boltzmann energy distribution with a characteristic energy $KT\ind{f}$. The photoelectron current (\ref{if}) is shown in magenta in Figure~\ref{fig:sweeps}.

As the lengths of the booms carrying the LAP probes are of the same order as the dimension of the spacecraft and the Debye length cannot be assumed to be much shorter, the electrostatic potential field from the spacecraft $\Phi(\vec{r})$ caused by the potential $V\ind{s}$ will not have decayed to zero at the probe position. We write its value as the location of the probe as $\Phi\ind{p}   = (1-\beta) V\ind{s}$, where $\beta$ is between 0 and 1. Whether photoelectrons are returned to the probe or not depends on the direction of the electric field at the probe surface. Therefore the relevant voltage in (\ref{if}) is the probe potential with respect to the local environment, $U\ind{p} = V\ind{p}-\Phi\ind{p}$, as verified in numerical simulations \cite{Johansson2016a} and laboratory experiments \cite{Wang2015a}.

\begin{figure}
\centering
\includegraphics[width=60mm]{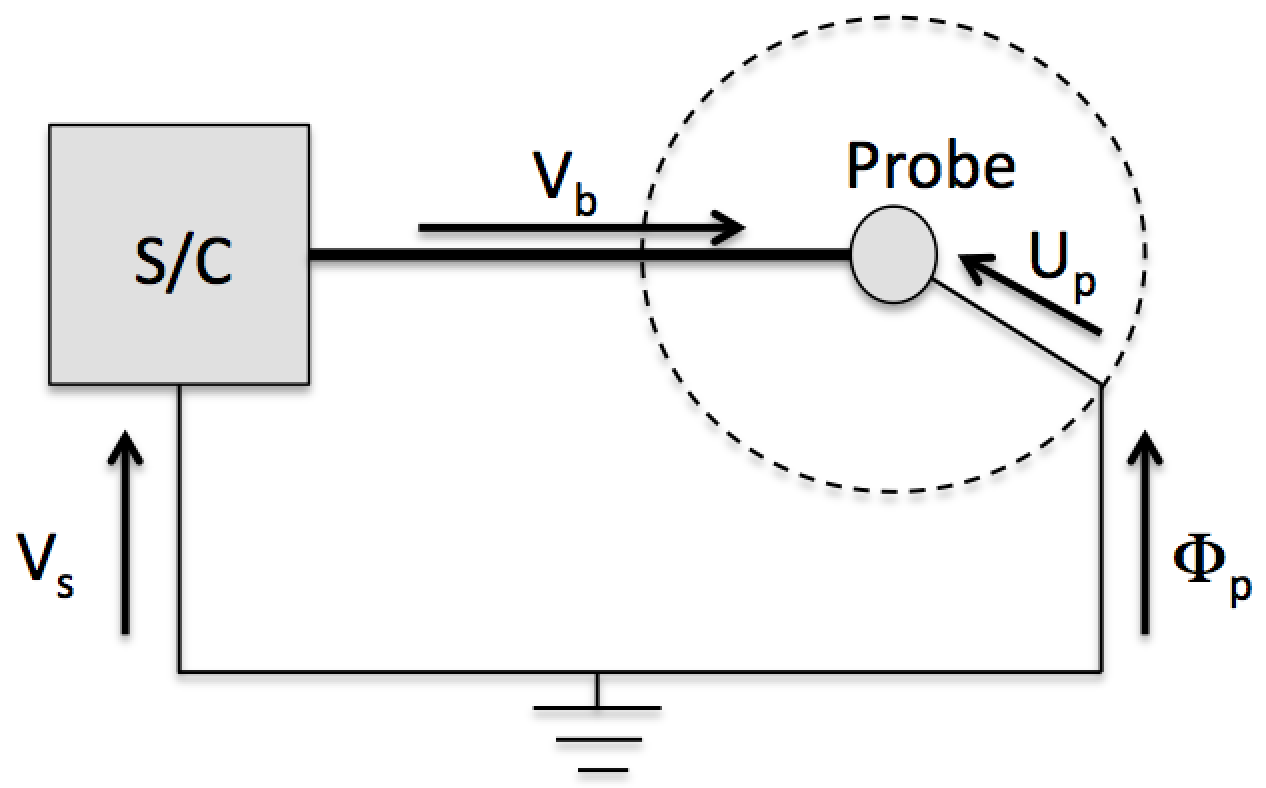}
\caption{Illustration of relevant potentials when there is no electric field in the unperturbed plasma. The bias voltage $V\ind{b}$ is set by the instrument, while the other potentials result from the interaction of spacecraft, probe, and plasma. The ground potential can be seen as a point far away, or as what the potential would have been at the location of the spacecraft had it not been there.}
\label{fig:skiss}
\end{figure}

The measured probe sweep will show a distinct signature when $U\ind{p} = 0$, that is,\ when the bias potential $V\ind{b}$ attains a value $V\ind{k} = \Phi\ind{p} - V\ind{s} = -\beta V\ind{s}$ (Figure~\ref{fig:skiss}). Comparisons of $V\ind{k}$ to the low-energy cutoff in ion energy observed by the Ion Composition Analyzer \citep[ICA;][]{Nilsson2015a}, caused by all ions having been accelerated through a potential drop $V\ind{s}$, show that $\beta \approx 0.8$ is typical for Rosetta in the inner coma during the main part of the mission \citep{Odelstad2016a,Odelstad2017a}. This means that $-V\ind{k}$ represents about 80\% of the spacecraft potential. This correction is minor for the purposes of the present paper, and we present observed values of $-V\ind{k}$ as estimates of $V\ind{s}$. 

We assume all ions are positive and singly charged, and that their thermal motion is small compared to the bulk speed in the spacecraft frame. Both assumptions should be reasonable as long as collisional coupling ties the low-energy ion bulk flow to the supersonic neutral gas \citep{Vigren2013a}. This may not be the case at the lowest activity level or far from the nucleus, but in these situations the ion current is small anyway. Based on COPS measurements for April-September 2015, \citet{Mandt2016a} estimated the ion collisional zone to extend outside of Rosetta's position for most of this interval. The ion current to a spherical probe at potential $V\ind{p}$ with respect to the unperturbed plasma far away is then 
\begin{align} \label{ii}
I\ind{i} = 
\begin{cases}
-I\ind{i0} \left(1 - \frac{eV\ind{p}}{E\ind{i} }\right), & V\ind{p} < E\ind{i}/e\\ 
0, & V\ind{p} > E\ind{i}/e,
\end{cases}
\end{align}
where the ram ion current at zero potential for a probe of area $A\ind{i}$ projected to the flow direction is
\begin{align}\label{ii0}
I\ind{i0} = n u e A\ind{i},
\end{align}
and the drift energy of ions of mass $m\ind{i}$ flowing at speed $u$ is $E\ind{i} = m\ind{i} u^2/2$. For a sphere, the area projected to the ion flow is $A\ind{i} = \pi a^2$. Equation~(\ref{ii}) can be seen as the cold ion limit of a more complete but complicated expression for the 
OML current in a flowing warm plasma \citep{Medicus1961a}. The full expression has the same linear dependence on $V\ind{p}$ for attractive potentials, and, as the ion current for $V\ind{p}>0$  is small compared to electron current, the main effect of violating the supersonic ion flow approximation is to change the interpretation of the ion speed $u$ and energy $E\ind{i}$. We note that in (\ref{ii}) we use $V\ind{p}$, the probe potential with respect to infinity, as for a negative $V\ind{s}$, no barriers will form between the probe and the plasma. It is not obvious that (\ref{ii}) should hold, given pre-acceleration of the ions through the potential drop $\Phi\ind{p}$ before reaching the vicinity of the probe, and realistic particle-in-cell simulations of LAP sweeps including the spacecraft indicate that the slope d$I\ind{i}/$d$V$ will indeed be smaller (Johansson et al., manuscript in preparation). This can be interpreted as the ions being picked up by LAP having a higher speed than the background flow due to acceleration through part of the spacecraft potential field. When fitting theoretical expressions to sweeps as in Figure~\ref{fig:sweeps}, where the ion current is shown in black, this would cause us to overestimate the ion bulk flow momentum, $m\ind{i}u$. 

In the OML formulation for a plasma at equilibrium, the electron current to an ideal isolated sphere at potential $V\ind{p}$ is 
\begin{align}\label{ie}
I\ind{e} = 
\begin{cases}
I\ind{e0} \exp \left(\frac{eV\ind{p}}{KT\ind{e}} \right), & V\ind{p} < 0\\ 
I\ind{e0} \left(1 + \frac{eV\ind{p}}{KT\ind{e}} \right), & V\ind{p} > 0,
\end{cases}
\end{align}
where the electron current due to random thermal motion is 
\begin{align}\label{ie0}
I\ind{e0} = A\ind{e} n e \sqrt{\frac{KT\ind{e}}{2 \pi m\ind{e}}},
\end{align}
and $A\ind{e} = 4 \pi a^2$ is the surface area of a spherical probe of radius $a$. In Figure~\ref{fig:sweeps}, the electron current is plotted for two electron populations of different temperatures, denoted warm (`w', light blue curve) and cold (`c', green).

For LAP on Rosetta, the direct application of this relation is complicated by part of the spacecraft's electrostatic field remaining around the probe. Unless the spacecraft potential field at probe position, $\Phi\ind{p}$, has decayed to well below $KT\ind{e}/e$, the electron distribution directly accessible to the probe will be perturbed. For Boltzmann electrons and a negative $V\ind{s}$, the number density in the neighbourhood of a point-like probe will be decreased by a factor $\exp(e\Phi\ind{p}/KT\ind{e})$. For a finite-sized probe at positive bias, the field from the probe itself alleviates this effect \citep{Laframboise1974a}. 

Adding the electrostatic field of a positively biased probe to the field from the negatively charged spacecraft means there may be no path from the probe to infinity along which the potential decays monotonically, causing a barrier to form for low-energy electrons from outside. Electron collection around the probe is then regulated not only by the probe potential with respect to the local plasma, $V\ind{p}$, but also by this  potential barrier. Theoretical considerations \citep{Olson2010a} indicate that (\ref{ie}) still holds in two limiting cases: The upper expression now applies for $U\ind{p} < 0$, that is,\ the limit is set by the potential with respect to the local plasma around the probe, while the lower expression applies when the bias exceeds a critical value $V\ind{c}$ needed for the probe to fully suppress the barrier and open a channel to the surrounding plasma. Finding $V\ind{c}$, and the current in between these voltages, are non-trivial tasks. \citet{Olson2010a} provided estimates using a simplified model, and also showed that the effect can be identified in data. This was also demonstrated in laboratory experiments by \citet{Wang2015a}, showing expected signatures both at $V\ind{b} = V\ind{k}$, where the probe is at the same potential as its immediate surroundings, and at $V\ind{b} =V\ind{c}$. Numerical simulations of LAP probe bias sweeps including the spacecraft (Johansson et al., manuscript in preparation) verify such effects, but also show that the cold ion current (\ref{ii}) retains its linear relation to voltage and goes to zero at the same value of $V\ind{p} = V\ind{s} + V\ind{b} = E\ind{i}/e$, thereby creating an opportunity for a robust estimate of $V\ind{s}$ for cases when $I\ind{i}$ can be reliably isolated from the other currents and $E\ind{i}/e$ is known or can be neglected. For the example sweeps in Figure~\ref{fig:sweeps}, we have not attempted to include barrier effects in the analysis, though we discuss them again in Section~\ref{sec:cold}.

\subsection{Tenuous regime ($n \lesssim 10^1$ {\rm cm}$^{-3}$)}\label{sec:tenuous}

For tenuous plasmas where the plasma electron flux is small, such as the unperturbed solar wind, the spacecraft attains a positive potential $V\ind{s}$ in order to retain a fraction of its photoemitted electrons sufficient for the total current to it to be zero \citep{Pedersen1995a}. This will be the case when the current carried by photoelectrons emitted from the spacecraft is higher than that of plasma electrons. Neglecting the difference between electron collection and photoemitting area on the spacecraft for an order of magnitude estimate, and using a photoemission current on the order of 10~$\mu$A/m$^2$ at 1~AU, we arrive  at a limiting value on the order of 5~cm$^{-3}$ for 3~AU, assuming $T\ind{e} \sim 10$~eV as typical of electrons in the solar wind as well as in the photoionised cometary coma. 

In this case, the ion current collected by a LAP sensor at negative voltage will be too small for reliable measurement, and the electron current will have a significant contribution from the cloud of photoelectrons surrounding the spacecraft because of its positive potential. An example sweep acquired far from the comet at 3.7~AU is shown in Figure~\ref{fig:sweeps}(a), together with the model expressions (\ref{if}) -- (\ref{ie}) with parameters as in Table~\ref{tab:fit}. Some of these parameters are well determined, while others, marked by parentheses, are not. To fit the photoemission current, $V\ind{k}$ must be close to the -1.1~V indicated, so $V\ind{s}$ is positive as is typical for spacecraft in the solar wind. As the Debye length expected in the solar wind is long, 10~m or more, there is little shielding between spacecraft and probe so $\beta$, the fraction of $V\ind{s}$ picked up by $V\ind{k}$, should be low. The actual value of $V\ind{s}$ may thus be several volts. The photoemission parameters $I\ind{f0}$ and $T\ind{f}$ are also well constrained. We note that we obtain a good fit by setting the temperature of collected electrons $T\ind{ew}$ to the same value as $T\ind{f}$, consistent with the expectation that the electron current to the probe is dominated by photoelectrons emitted by the spacecraft; $n\ind{ew}$ may thus be interpreted as the density of the photoelectron cloud near the position of the probe. While the model currents are drawn with the ion parameters shown, these parameters are in fact not constrained at all by the measured sweep as the ion current is too small to be detectable for any realistic solar wind parameters.  

While the plasma density $n$ in this regime cannot be determined by fitting the ion and electron currents, it can be estimated from $V\ind{s}$ \citep{Pedersen1995a}, typically by use of an empirical fit to some other density measurement. This technique was used for Rosetta by \citet{Edberg2009a} at the Mars swing-by, and at 67P by \citet{Odelstad2015a}.

\subsection{Intermediate regime ($10^1$ {\rm cm}$^{-3} \lesssim n \lesssim 10^3$ {\rm cm}$^{-3}$)}\label{sec:intermediate}

For intermediate density plasmas, which we take to mean from a few tens to a few thousands of particles per cm$^3$ with electron temperature $T\ind{e}$ on the order of a few to about 10 eV, the plasma electron flux overcomes the spacecraft photoemission and the ion flux resulting in $V\ind{s}<0$. This is the most typical environment seen by Rosetta at the comet \citep{Edberg2015a,Odelstad2015a}. In this situation, there are several means to access the plasma parameters using the Langmuir probes:

\begin{enumerate}
\item
As the electron current is now large, the Langmuir probe current-voltage characteristic can be directly used to determine the local electron density $n$ and temperature $T\ind{e}$ using relevant theoretical models (\ref{fig:sweeps}). Spacecraft photoelectrons are driven away by the negative spacecraft potential, but could still in principle contribute to the probe current. However, due to spacecraft operational requirements and pointing constraints, the LAP probes were almost always behind the plane defined by the solar panels, even when sunlit (on the `night side' of the spacecraft `terminator plane', i.e.\ the Sun is in the +X direction in Figure~\ref{fig:rpc}, with the solar panels appropriately tilted to face the Sun). For positive spacecraft potentials (Section~\ref{sec:tenuous}), the photoelectron trajectories are bent around the spacecraft by the attractive spacecraft electric field and can thus be picked up by the probes, as was seen in the example in Figure~\ref{fig:sweeps}(a). For negative potentials, the photoelectrons go away more or less radially, therefore not reaching the probes.     

The Debye length in this parameter range is still much greater than the size of the LAP probes, so variants of the orbital motion limited (OML) theory \citep{Mott-Smith1926a,Medicus1961a,Laframboise1973a,Wahlund2005a} can be used for data interpretation (Figure~\ref{fig:sweeps}(b) and (c)). 
However, for typical $T\ind{e}$ values of $\sim 5$ eV, the boom length is still not very much shorter than the Debye length, which must be kept in mind in the interpretation of data \citep{Laframboise1974a,Olson2010a,Wang2015a}. In the limit where the plasma electron flux greatly exceeds the flux of electrons from the spacecraft by photoemission and secondary emission, $V\ind{s}$ is expected to approach a negative value several times the electron thermal energy equivalent. As the boom length must still not be shorter than the Debye length, we may expect the fraction of $V\ind{s}$ picked up by the probes to stay below unity, which has been confirmed by comparison of LAP photoemission data to the lowest ion energy visible in ICA data \citep{Odelstad2016a,Odelstad2017a}. In this situation, it may also be necessary to consider that the local electron density around the probe is reduced by a factor $\exp(e \Phi\ind{p}/KT\ind{e})$, where $\Phi\ind{p}$ is the potential in the spacecraft sheath at probe position \citep{Odelstad2015a}. For Boltzmann distributed electrons, this will not impact on the estimate of $T\ind{e}$ based on a fit to the exponential part of the Langmuir probe characteristic. The exponential relation between density and total energy for a repelling potential means the effect of a shift in potential is just a numerical factor not changing the shape of the energy distribution.
\item
The ion current to the probe when at negative bias voltage grows significant in plasmas of intermediate density, and so can be used to measure the density. In the terrestrial ionosphere, this technique is frequently used, preferentially with planar probes facing the ram direction, therefore
 the ion current only depends on the density as long as the ion motion is supersonic in the spacecraft frame \citep{Brace1998a}. Spherical probes have also been similarly used, for example on Cassini in the Saturn plasma torus \citep{Holmberg2012a}, but in this case the current also depends on ion mass \citep{Wahlund2005a}. For Rosetta in the inner coma of comet 67P, the ion flow may still be assumed to be supersonic, as the ion bulk flow speed as well as temperature will be set by collisional coupling to the neutral gas, which is supersonic due to its expansion into what is essentially a vacuum \citep{Tenishev2008a,Combi2012a}. However, in contrast to the ionospheric cases mentioned, the known motion of the Rosetta spacecraft in the reference frame of the main body is small compared to the a priori unknown flow speed of the ions, meaning either the gas speed or the effective ion mass (the harmonic mean of all species) has to be supplied from models or other measurements. While the functional form of the OML expressions for the current-voltage relation formally allows fitting density, effective ion mass, flow speed and spacecraft potential all at the same time, errors easily grow large in this process, particularly if emission of photoelectrons and secondary electrons also enter the problem and have to be corrected for \citep{Garnier2012a,Holmberg2012a}. Additional information on some parameters is therefore always desirable and often necessary.
\item
Alternatively, the plasma density $n$ can be estimated from $V\ind{s}$ also in this regime \citep{Odelstad2015a}. This does not work when only currents proportional to the plasma density, that is,\ currents due to collection of plasma ions and electrons, carry charge to the spacecraft, but as long as the spacecraft photoelectron emission current is not negligible compared to the ion current, $V\ind{s}$ will depend on $n$ as well as on $T\ind{e}$. This makes it possible to estimate $n$ from $V\ind{s}$. 
An advantage of this method is the high consistency achievable in the $V\ind{s}$ estimate; though the possibly varying $T\ind{e}$ as well as any effects of secondary electron emission add uncertainty in this regime.  
\end{enumerate}

Two sweeps from this regime are shown in Figure~\ref{fig:sweeps}(b) and (c). While the sweep in Figure~\ref{fig:sweeps}(b), obtained at low comet activity, can be fitted with one single electron population, this is not possible for its counterpart in a denser plasma in Figure~\ref{fig:sweeps}(c) where two electron populations at different temperatures are needed for combining the high slope at the right with the extended exponential-like decay to the left of the steep part. 

In the intermediate regime, the Debye length is not short compared to the boom length. For the fit parameters to the sweep in Figure~\ref{fig:sweeps}(c), tabulated in Table~\ref{tab:fit}, we get $\lambda_D \approx 1$~m, implying a significant fraction of the spacecraft potential may remain in the surroundings of the LAP probes. {\  The fitted temperature of the cold electrons is only 1~eV and the s/c potential from $-V\ind{k}$ is at least -18.9~V, which should not have decayed to zero at the probe position. Therefore, it is not obvious how electrons of energy 1~eV are able to reach the probe even when taking into account the modification of the potential structure in space by the probe itself as discussed at the end of Section~\ref{sec:intermediate}. On the other hand, the sweep clearly shows a steeper slope at high positive voltages than can be explained by the warm population alone. This discrepancy points to the energy distribution being more complex than what our simple two-population model can handle, with a higher fraction of the electrons at intermediate (few eV) energy than the model allows. Such non-Maxwellian distributions are to be expected in a marginally collisional plasma, and it should be no surprise that our simple model cannot reproduce all details. Nevertheless, sweeps of the type in Figure~\ref{fig:sweeps}(c) need a mix of electrons that have and have not lost energy by collisions on the neutral background, even if the model of two distinct Maxwellian populations cannot handle this properly.} 

There could also be times when a cold population is entirely invisible in LAP electrons through the barrier effect discussed above (after Equation~\ref{ie}). As the fit is made both to the electron and ion sides and the fit value for the density is thus constrained by the ions, the main issue presumably is with $T\ind{ec}$, which could be highly exaggerated. Awaiting systematic cross-calibrations with the MIP instrument, one should regard cold electron parameters from LAP sweep fits as relatively uncertain. Barriers will only affect electrons, so the ions do not have this problem. However, as noted above in the discussion of Equation~\ref{ii}, the slope on the ion side may be underestimated in these circumstances, making us overestimate ion momentum. The flow speed value of 2~km/s is therefore badly constrained and should not be taken at face value, which is why we put it in parentheses in Table~\ref{tab:fit}. 

\subsection{Dense regime ($n \gtrsim 10^3$ {\rm cm}$^{-3}$)}\label{sec:dense}

For the highest density plasmas in the inner coma, $T\ind{e}$ is expected to fall to values typical of the neutral gas, which can be a few hundred kelvin or even lower \citep{Tenishev2008a}. This corresponds to a few per cent of an eV, which is outside the practical limit of what a Langmuir probe can resolve from the exponential part of the I-V curve, limited not only by variations of the work function over the probe surface and the smallest bias voltage step available \citep{Eriksson2007d} but in reality also by temporal variations of the plasma and  the spacecraft potential. An example of such an I-V curve is shown in Figure~\ref{fig:sweeps}(d). To fit this sweep, two electron populations have been used, one cold, to explain the steep slope on the right part of the curve, and one warm, for explaining the negative $V\ind{s}$ \citep{Johansson2016a} as well as for fitting a small observable exponential-like fall-off at voltages just to the left of the steep part. Due to the presence of the warm population as well as by the limitations imposed by the LAP bias voltage step of 0.25~V used here, the temperature $T\ind{ec}$ of the cold population cannot be resolved directly, but testing different values shows it could not have exceeded 0.1~eV without being visible in the sweep. 

In the dense regime, the Debye length is short compared to the boom length. For the fit parameters to the sweep in Figure~\ref{fig:sweeps}(d), we get $\lambda_D \approx 5.5$~cm. Comparing to the intermediate regime in Section~\ref{sec:intermediate} above, we expect the cold electron parameters fitted to sweeps in dense plasmas to better represent the real plasma. Nevertheless, cross-calibration with MIP will be needed to constrain the numerical accuracy of values derived from LAP.

\begin{figure*}
\centering
\includegraphics[width=15cm]{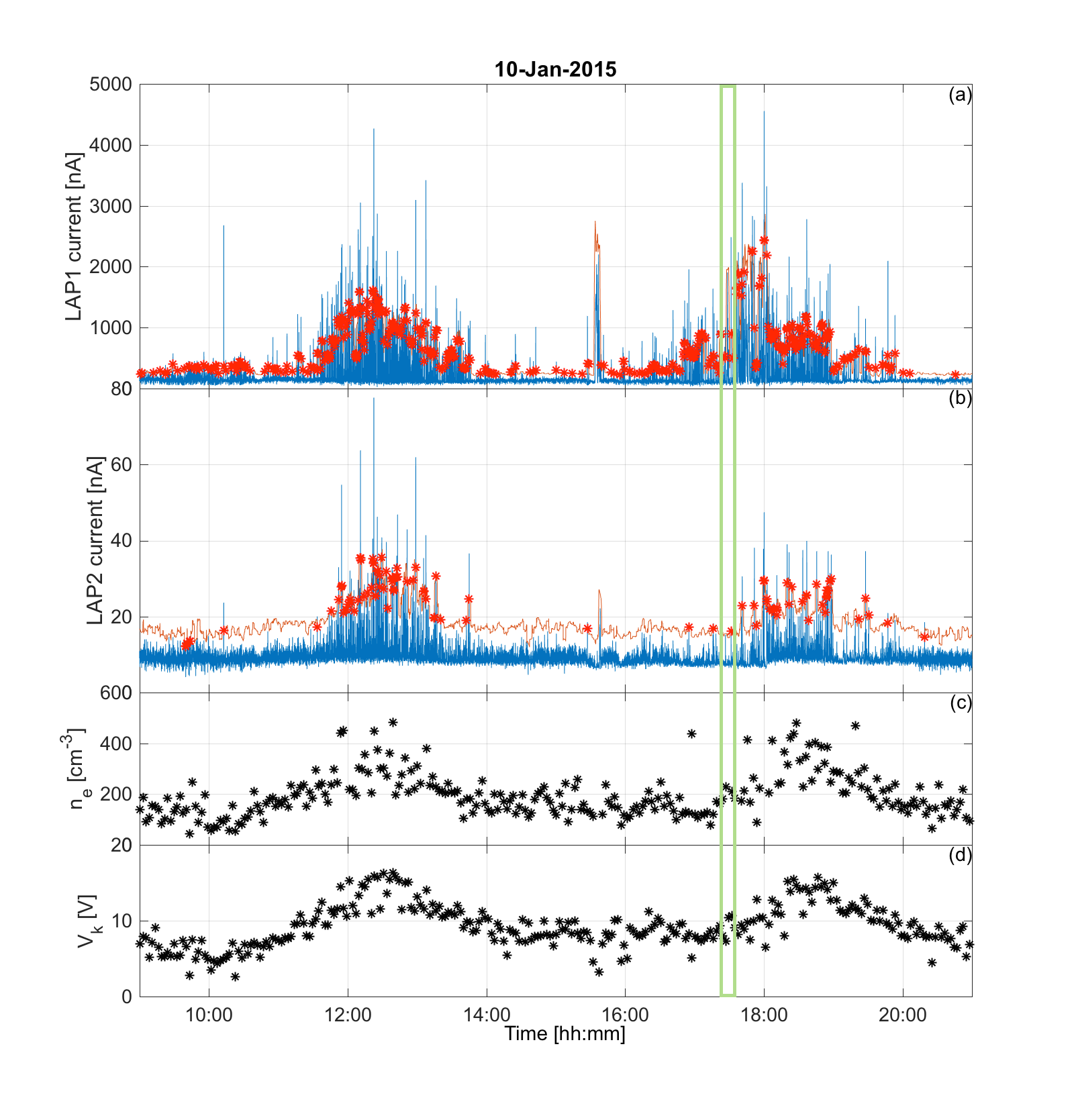}\\
\caption{Example of pulses in LAP1 (at +30~V, panel~(a)) and LAP2 (at -30~V, panel~(b)) current  (blue). In red is shown the dynamic threshold used to define pulses, with red dots marking the pulses found by the algorithm. Panels~(c) and (d) show sweep fit results for electron density and the negative of the spacecraft potential. The sign of the LAP2 current has been inverted, to show density increases as positive. The green box marks the interval zoomed into in Figure~\ref{fig:zoom}.}
\label{fig:pulse}
\end{figure*}

\section{Observations}\label{sec:cold}

The two sweeps presented in Figure~\ref{fig:sweeps}(b) and (c) showed obvious signatures of an electron population at typical temperature of 5-10~eV, referred to here as "warm". That this population is present more or less all the time since Rosetta started its near nucleus operations in September 2014 is clear from the mapping and statistics of the spacecraft potential by \citet{Odelstad2015a} and \citet{Odelstad2017a}, since electrons must have an energy of order $-e V\ind{s}$ to drive the spacecraft to a negative potential $V\ind{s}$. What determines $V\ind{s}$ is the electron flux, not the density, so a negative $V\ind{s}$ cannot be taken as direct evidence of warm electrons dominating the density, and indeed the fit in Figure~\ref{fig:sweeps}(c) needs a higher density for cold electrons (Section~\ref{sec:cold}) than for the warm population.



As discussed above, the neutral gas density falls off roughly as $1/r^2$ and stays cold. Electrons released by photoionisation near the nucleus may see sufficiently high neutral density to cool by collisions on the cold neutrals. While Rosetta is mainly outside the region of such cooling \citep{Mandt2016a}, it is possible that when such a collisional region is present somewhere closer to the nucleus along the flow line of gas expansion connecting the nucleus to Rosetta, we could observe a remnant cold electron population in addition to the warm photoelectrons created outside this region, unless efficient heating processes would destroy it.

Figures~\ref{fig:sweeps}(c) and (d) show examples of LAP probe bias sweep where cold electrons are needed to fit the data. In particular, steep sweeps with no visible exponential region, like the example in Figure~\ref{fig:sweeps}(d), present clear evidence of an electron population with $T\ind{e} \lesssim 0.1$~eV. It is clear that this cold population cannot be the only electrons present, since the highly negative spacecraft potential would otherwise have been impossible to maintain. This agrees with the expectation of two electron components at different temperatures. The electron current reached in this sweep, 15~$\mu$A, is among the highest observed by LAP during the whole mission, mostly due to the low temperature of the cold electron population. These data were obtained in the highest-activity phase near perihelion, and are taken from near a diamagnetic cavity crossing \citep{Goetz2016a,Goetz2016b} when such cold plasma could be observed for extended intervals up to several hours. This sweep contrasts the example in Figure~\ref{fig:sweeps}(b) taken early in the mission, at low neutral gas density where no efficient cooling of electrons can be expected, for which we did not need to introduce any cool electron population to fit the data.

While statistics of cold plasma observations have to wait for another study, we note that extended intervals showing signatures of dense cold plasma, similar to the example sweep in Figure~\ref{fig:sweeps}(d), were mainly found around perihelion. However, large currents of cold electrons first turned up in the LAP measurements as short pulses in the probe current, and have been seen in this
shape during most of the mission. As these signatures are very conspicuous in LAP data, we discuss them here.

\begin{figure}
\centering
\includegraphics[width=\columnwidth]{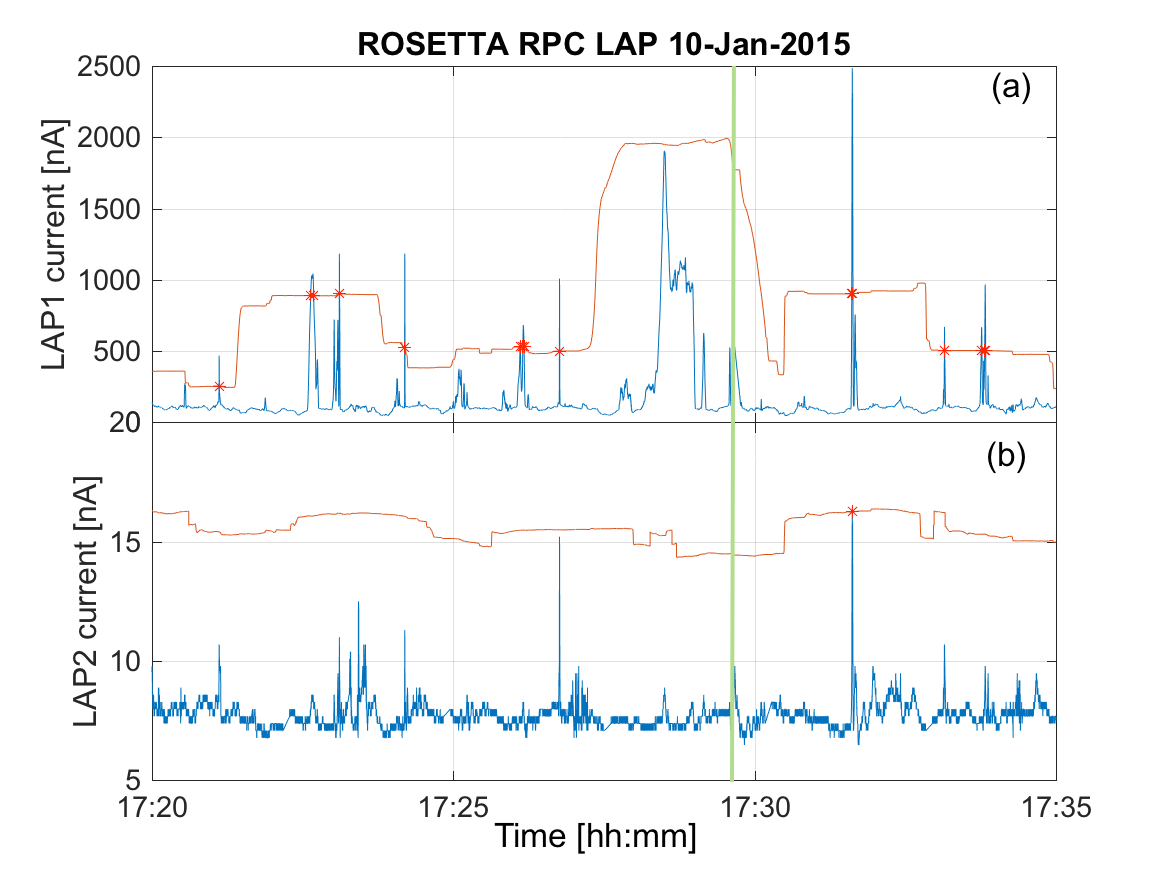}
\caption{Zoom in to part of the interval in Figure~\ref{fig:pulse} to show details of pulses. The current here is sampled at 57.8~Hz, but very little structure is to be seen at timescales shorter than about a second. Blue is measured current and red is the dynamic threshold for pulse detection defined in the text. The vertical green line indicates the time of the sweep in Figure~\ref{fig:swp-pulse}.}
\label{fig:zoom}
\end{figure}

An example interval with pulses in the probe current at fixed bias potential is seen in Figure~\ref{fig:pulse}, with a zoom-in to 15~minutes of the data in Figure~\ref{fig:zoom}. The blue curves in Panel~(a) of both Figures show the current to LAP1 in this interval at a bias potential $V\ind{b} = +30$~V with respect to the spacecraft in between the short (3~s at this event) probe bias sweeps (occurring every 160 s). As the spacecraft potential $V\ind{s} \approx -V\ind{k}$ at the time varies between $-5$~V and $-15$~V, the potential of LAP1 with respect to the plasma $V\ind{p} = V\ind{s} + V\ind{b} \gtrsim 15$~V (Figure~\ref{fig:skiss}) is sufficient for attracting electrons and repelling ions. Panel~(b) shows the negative of the current to LAP2. At a bias of $-30$~V, LAP2 is around 40~V negative with respect to the plasma, and as the photoemission can be assumed to be stable on this time scale, any variations in the probe current should be due to varying ion flux. The two lower panels in Figure~\ref{fig:pulse} show results from sweep fits. The plasma density in Panel~(c) derived from LAP sweeps using the slope of the probe curve at positive potentials (second expression in Equation~\ref{ie}) assuming $T\ind{e} = 5$~eV, and  (Panel~(d)) the spacecraft potential proxy $-V\ind{k}$. More details on the background plasma in this event, including detailed modelling of the ionisation, can be found in \citet{Vigren2016a}.

The background value of the electron current to LAP1 (Figure~\ref{fig:pulse}(a)) is around 100--200~nA, over which rise huge pulses of several $\mu$A. Figure~\ref{fig:zoom}(a) shows pulses extending typically from a few to several tens of seconds. The large pulse between 17:28 and 17:29 thus carries around 50~$\mu$C$ \approx 3 \cdot10^{12} e$ to the probe. We also find pulses in the ion current simultaneously observed at the negatively biased LAP2, but at lower magnitude both in absolute sense and relative to the background current (from which we have subtracted a constant value of 5~nA to account for probe photoemission). Some pulses coincide well in both signals, while for example the very large electron current pulse discussed immediately above has only a weak signature in the ions. Comparing panels~(a) and (b) to panel~(c) shows more frequent and higher pulses where the density is high.



Simultaneous detection on probes sampling electrons and ions indicates a local increase in plasma density as the source of a pulse. The generally higher increase in electron current compared to ion current suggests the increase is mainly due to electrons of very low temperature (c.f.\ Equation~\ref{ie}). Examples such as the large electron current pulse 17:28-17:29 in Figure~\ref{fig:zoom}, for which the ion current increase is weak, could possibly be due to only the local electron temperature decreasing without any density increase, as this would also increase the electron current (Equation~\ref{ie}). However, some caution is needed in the interpretation of individual structures as the ions are flowing and small changes in direction due to electric fields related to any plasma process may cause local wake effects and blocking of the ion flow to the probe by spacecraft structures. Further discussion of the interpretation of the pulses follows in Section~\ref{sec:filament}.


\begin{figure}
\centering
\includegraphics[width=6 cm]{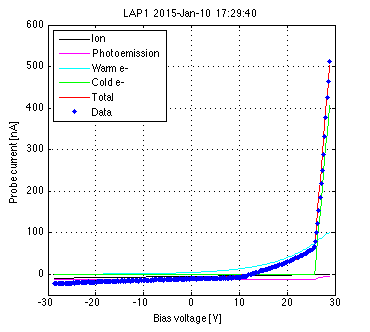}
\caption{Probe bias sweep obtained inside a pulse at the time marked in Figure~\ref{fig:zoom}. The format is similar to Figure~\ref{fig:sweeps}.}
\label{fig:swp-pulse}
\end{figure}

The probe bias sweeps sometimes coincide with a pulse. Figure~\ref{fig:swp-pulse} shows a typical example. Fit parameters for this sweep are tabulated in the right column of Table~\ref{tab:fit}, with parenthesis indicating those that we judge to be highly uncertain. The fit is quite good, but there is still strong reason to distrust many of the parameters. The presence of two clear knees in the sweep, at about 12~V and 25~V, suggests the barrier effects discussed in Section~\ref{sec:interp} are important here \citep{Olson2010a,Wang2015a}, due to the negatively charged spacecraft repelling low-energy electrons from its vicinity. However, the high slope seen above 25~V can only be due to collection of a cold electron population. This is because, regarding Equation~\ref{ie}, this slope is proportional to $n/\sqrt{T\ind{e}}$, but the density value is constrained by the slope due to ion collection at negative voltages, which is proportional to $n/(m\ind{i} u)$. As noted above, the sweep likely underestimates the slope on the ion side, because the speed of ions collected by the probe is higher than the unperturbed flow speed $u$ by acceleration of the ions towards the negative spacecraft. This can explain the high ion flow speed (about four times above neutral gas speed) used in the fit, though we note that recent modelling indicates that even weak electric fields can accelerate some ions to about the values we get here, despite the presence of collisions \citep{Vigren2017a}. Nevertheless, if we try to explain the high slope on the electron side with a high density of warm electrons only, putting the density $n\ind{ec}$ to zero, we have to increase the already high ion momentum to much higher numbers in order to keep the ion current at observed values. The fitted values of $n\ind{ec}$ and $T\ind{ec}$ are both numerically uncertain and we do not claim we have an observation of 0.02~eV electrons (200~K), but sweeps like this demonstrate the presence of an electron population with $T\ind{ec}$ below 0.1~eV in the pulses. 

To investigate where the pulses of cold plasma are seen, we used a simple algorithm for their detection and ran on all LAP1 data for 2015. The algorithm works by defining a threshold current from a 128-second interval of data and considering LAP currents above this threshold to be pulses. If $I\ind{min}$ is the minimum value of the current in that interval, the threshold is set as  $\max(I\ind{min}+100~\mathrm{nA}, 2\, I\ind{min})$ for the electron side and $\max(I\ind{min}+2~\mathrm{nA}, 2\, I\ind{min})$ for the ion side. These values were chosen to avoid false detections in noisy environments, at the cost of low efficiency in detection. An example of how the algorithm works can be seen in Figure~\ref{fig:zoom}. The red curves give the dynamic threshold value, with the stars indicating identified pulses in LAP1 and LAP2 currents. It is clear that many of the pulses are missed, including a large and wide one in LAP1 around 17:29. It is also clear that more pulses are detected in the electron current than in the ion current, but this in large part an effect of their different amplitudes on the detection algorithm, which favours large pulses. However, no false detections are seen. Work on improving the algorithm for better statistics is ongoing, but this version works for a first idea of where pulses occur.

\begin{figure*}
\centering
\includegraphics[width=18cm]{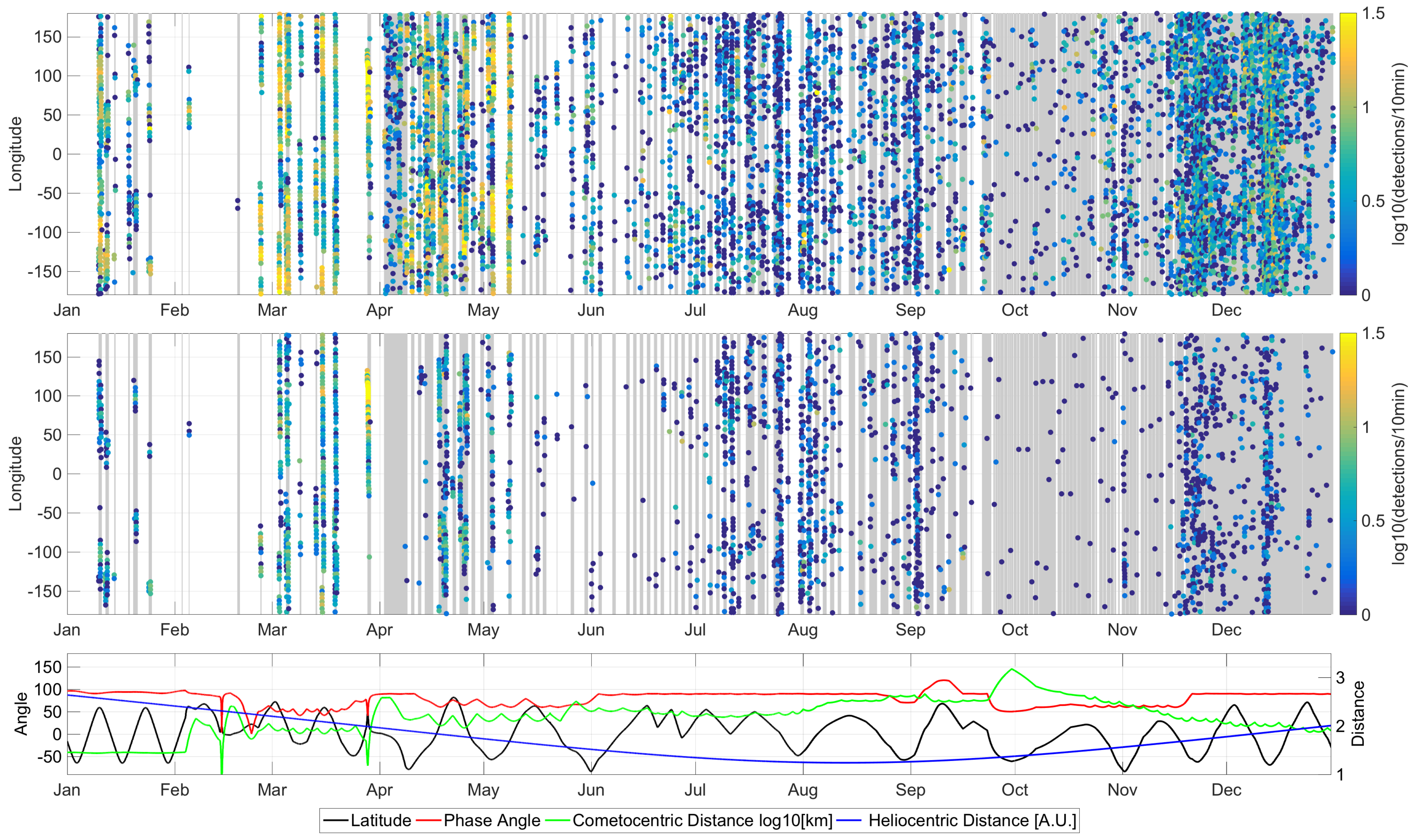}
\caption{Distribution in time and space of current pulses in LAP, based on the simple algorithm described in the text. Grey indicates regions searched where no pulses were detected by the algorithm; white periods indicate regions without suitable data. {\bf Top:} Number of pulses detected in LAP1 probe current per 10-minute period, corresponding to $5^\circ$ cometary longitude, during 2015. {\bf Centre:} Number of large pulses per 10-minute period. {\bf Below:} Latitude and phase angle (solar zenith angle) on left scale, heliocentric distance in AU and $^{10}$log of cometocentric distance in km on right scale.}
\label{fig:odel}
\end{figure*}

A first overview of the results of running the pulse-finding algorithm, such as it is, on the LAP1 data for 2015 is given in Figure~\ref{fig:odel}. The upper panel shows the number of pulses detected per 10-minute interval, plotted versus time and longitude of the sub-spacecraft point on the nucleus. Each nucleus rotation of around 12~hours is thus a vertical stripe in the plot, and each data point of ten minutes corresponds to roughly 5$^\circ$ in longitude on the nucleus. Additional orbital information is given in the lower panel. In the upper panel, grey areas mark all time intervals examined (which are all when LAP1 was in a mode suitable for pulse detection) where no pulses were found. Intervals not searched are white. The colour scale is logarithmic, so the number of detected pulses per ten-minute interval varies here between 1 and 30. To give an idea of where the largest pulses occur, the centre panel shows the same kind of data as the upper though restricted to pulses of high amplitude, achieved by changing the algorithm thresholds above by replacing the offsets 100~nA and 2~nA by 1~$\mu$A and 20~nA, respectively. Despite the shortcomings of the finding algorithm, comparisons to sample data indicate that Figure~\ref{fig:odel} provides a reasonable view of where pulses occur, though the actual values given for the number of pulses per ten-minute interval are likely underestimated.

The first thing to note is that while pulses are detected throughout the mission, their distribution is not uniform. Comparing the statistics in the upper panel to the orbit information in the lower, we can note that when Rosetta is in the northern hemisphere (positive latitudes, black curve), detections cluster around longitudes $\pm 100^\circ$, corresponding to being above the neck region, at least during northern summer (up to May). This reflects similar behaviour to that known for the neutral gas density \citep{Hassig2015a,Bieler2015a}, plasma \citep{Edberg2015a,Odelstad2015a} and dust \citep{Rotundi2015a,Fulle2015a}. From early May, the sun was in the southern hemisphere of 67P, which therefore picked up activity \citep{Hansen2016a}, and at least by early June we find most pulses at negative latitudes. In July-September, large pulses are detected mostly at low latitudes, and become a substantial fraction of the total in the southern hemisphere. During the dayside excursion in late September, the number of pulses decreases with cometocentric distance, recovering only slowly as Rosetta returns toward the nucleus in October. Only when back within about 150~km distance in mid November do the numbers really pick up. There may possibly also be a phase angle effect, favouring terminator over dayside, though disentangling this from distance will require a more detailed analysis. What is very clear is that when being back near the nucleus from mid November, most pulses are seen in the southern hemisphere, generally more active at that time. This is particularly clear for the highest-amplitude pulses (centre panel).

To a large extent, the pulses thus occur where the plasma density is highest. We discuss the pulses further in Section~\ref{sec:filament} below.

\section{Interpretation}\label{sec:disc}

\subsection{Warm electrons}

The warm electron population found in the sweeps in Figure~\ref{fig:sweeps} and inferred from the negative spacecraft potential \citep{Odelstad2015a,Odelstad2017a} is also detected by the RPC-IES electron spectrometer. \citet{Broiles2016a} studied this by fitting IES energy spectra to kappa distributions at two helicentric distances, 3~AU and 1.3~AU. At 3~AU, they found typical densities of 10--30~cm$^{-3}$ and temperatures of 10--30~eV. At 1.3~AU, corresponding ranges were 10--100~cm$^{-3}$ and 5--10~eV.  This is generally consistent with LAP densities for the warm electrons (see examples in Table~\ref{tab:fit}), though much of the density at perihelion can sometimes reside in a cold population invisible to IES (Figure~\ref{fig:sweeps}d and Section~\ref{sec:cold}). The temperature ranges are broadly consistent, though somewhat higher in IES than what we typically derive from LAP sweeps, in the examples and elsewhere. As the Langmuir probes only measure total current, any fit of the electron current, such as those shown in Figure~\ref{fig:sweeps}, will be more sensitive to the lowest energies than to the higher ones, while the case is opposite for IES, being directly mounted on the negatively charged spacecraft and hence only being exposed to a fraction of the low-energy electrons. The differing temperature estimates could thus in part be due to a combination of the two methods picking up characteristics of different parts of an energy distribution which is not Maxwellian, and in part due to effects of the repelling spacecraft potential. While at least the upper end of the $\kappa$ range found by \citet{Broiles2016a} suggests quite Maxwellian behaviour, at the lowest energies a kappa model may not well describe the actual electron distribution, as is clear from the presence of a cold population in case (c). 


\subsection{Electron cooling}\label{sec:colle}

As discussed in the Introduction, electrons released by photoionisation are expected to have a typical energy in the 12--15~eV range, resulting in $T\ind{e} \sim 10$~eV if no cooling occurs. This is also sufficient for explaining the observed negative spacecraft potentials of the same order, as a simple model of the equilibrium potential of an object in a thermal plasma predicts negative potentials of a few times $KT\ind{e}/e$, depending on the ion distribution and detailed object geometry. As noted by \citet{Mandt2016a}, Rosetta is usually outside of the region where strong cooling on the neutrals is expected, so observing this warm population is expected. Wave heating processes, invoked by \citet{Broiles2016b} to explain elevated fluxes of higher energy electrons (hundreds of eV) may also play a role, but the observed electrons in the few to ten eV range can be explained as excess energy from ionisation. In a sufficiently dense environment, at least some of the electrons will transfer energy to the neutral gas by various collisional processes \citep{Vigren2013a} and thus be cooled to as low as the few hundred kelvin (or even lower) expected for the expanding neutral gas. This requires high collisionality and thus high neutral gas density, as the mean free path is 
\begin{align}\label{lambda}
\lambda = 1/(n\ind{n} \sigma\ind{en}), 
\end{align}
where $n\ind{n}$ is the neutral gas density and $\sigma\ind{en}$ is the electron-neutral cross-section. Taking $n\ind{n}$ to decrease as $1/r^2$ with cometocentric distance from a value $n_0$ at the nucleus surface $r = R$, 
\begin{align}\label{n(r)}
n_n r^2 = n_0 R^2,
\end{align}
and the collision length will increase as $r^2$. Therefore, even if collisionality is high close to the nucleus, it will, for any given gas parcel, expand radially and become negligible at some distance. Observation of a mix of warm electrons (products of recent ionisation) and cold electrons (having lost energy to the neutrals) is therefore expected.

\citet{Mandt2016a} set the limit for electron collisionality, known as the electron collisionopause or exobase, as the cometocentric distance $r\ind{ce}$ at which 
\begin{align}\label{rce1}
\lambda(r\ind{ce}) = r\ind{ce,}
\end{align}
that is,\ at the point where the distance equals the collision length. By combining expressions (\ref{lambda}) to (\ref{rce1}) we find that
\begin{align}\label{rce}
r\ind{ce} = \sigma\ind{ne} n_0 R^2 = \sigma\ind{ne} n_n r^2.
\end{align} 
\citet{Mandt2016a} then used the daily average of the COPS neutral density data to plot this distance and compare it to Rosetta position for the period April-September 2015 (Fig.~5 of that paper). While neutral density appeared to be sufficient for this boundary to exist during all this period, Rosetta was found to stay outside of it all the time. A two-temperature electron gas is thus not unexpected, particularly when $r\ind{ce}$ is not too large, as the cooling boundary should be sharper the smaller the scale length in the density gradient at the position of the boundary. 

As a first estimate of the relative abundance of the two electron populations, we use the same assumption of a cooling boundary and combine this with a modified Haser model. For a plasma originating from and being coupled to a neutral gas expanding radially at constant speed $u$ from a spherical comet nucleus of radius $R$, the neutral gas density follows \citep{Haser1957a}
\begin{align}\label{nn(r)}
n_n(r) = n_0 \left(\frac{R}{r}\right)^2 \exp\left(\frac{\nu}{u}[R-r]\right) \approx n_0 \left(\frac{R}{r}\right)^2,
\end{align}
where $n_0$ is the density at the nucleus and $\nu$ is a constant ionisation frequency, and the last step follows by neglecting higher order terms in $\nu r/u$ in a Taylor expansion of the exponential. This is certainly valid in the inner coma as the ionisation timescale is at least $10^5$~s \citep{Vigren2015b,Galand2016a} and the flow speed $u$ is of order km/s. As the exponential loss describes the ionisation, the plasma density is given by the missing neutrals
\begin{align}\label{ne(r)}
n\ind{e}(r) & = & n_0 \left(\frac{R}{r}\right)^2 \left[1 - \exp\left(\frac{\nu}{u}[R-r]\right)\right] \nonumber \\ 
 &\approx& n_0 \frac{\nu R}{u} \frac{R}{r} \left(1 - \frac{R}{r}\right),
\end{align}
where we used the same series expansion. We now modify this by considering a cooling boundary $r\ind{ce}$, and assume we are several cometary radii from the nucleus and so can neglect the last term in (\ref{ne(r)}). Inside $r\ind{ce}$, all electrons are assumed cold, and their density will follow 
\begin{align}\label{nc(r)}
n\ind{c}(r) \approx n_0 \frac{\nu R}{u} \cdot \frac{R}{r}, & &  r \leq r\ind{ce}.
\end{align}
As no new cold electrons appear outside $r\ind{ce}$ we must in this region have
\begin{align}\label{nc2(r)}
n\ind{c}(r) = n\ind{e}(r\ind{ce}) \frac{r\ind{ce}^2}{r^2} = n_0 \frac{\nu R}{u} \cdot \frac{R r\ind{ce}}{r^2}, &  & r > r\ind{ce}.
\end{align}
An indicative relation for the fraction of cold electrons outside the cooling boundary can be obtained by dividing (\ref{nc2(r)}) by (\ref{ne(r)}), which with the above approximations simply yields an inverse distance law,
\begin{align}\label{coolfrac}
\frac{n_c}{n_e} = 
\begin{cases}
1, & r \leq r\ind{ce}\\
r\ind{ce}/r, & r > r\ind{ce}.
\end{cases}
\end{align}
The result is independent of $u$ and $\nu$ but depends on the value of $\sigma\ind{en}$ through $r\ind{ce}$. We have used $\sigma\ind{en} = 1.5 \cdot 10^{-19}$~m$^2$, taken from Fig.~2 of the compilation by \citet{Itikawa2005a} as relevant for 3--20~eV electrons impacting on water molecules.

\begin{figure}
\centering
\includegraphics[width=\columnwidth]{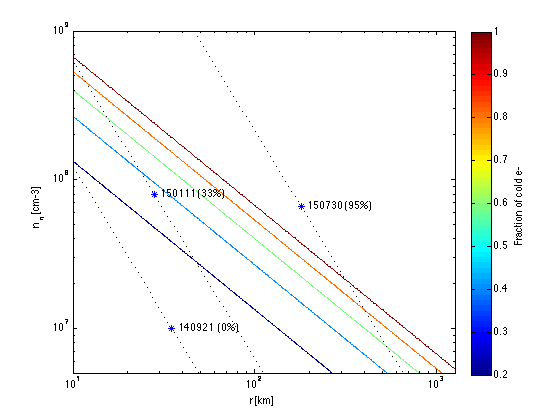}
\caption{Expected fraction of cold electron gas as a function of radial distance and neutral density - contour plots for 20\% (marine), 40\% (blue), 60\% (green), 80\% (orange), and 100\% (brown). The black dotted lines show $1/r^2$ neutral density profiles for the COPS values observed at the points marked with stars on the dates shown. The fraction of cold electrons in the sweep fits in Table~\ref{tab:fit} is indicated in parenthesis.}
\label{fig:nn}
\end{figure}

Figure~\ref{fig:nn} shows contours of constant $n_c/n_e$ in the $r$--$n\ind{n}$ plane calculated from (\ref{coolfrac}). At any given cometocentric distance $r$, increasing neutral density means a higher fraction of cold electrons. The figure also includes three data points representing the neutral density obtained from COPS measurements close to the times of the three LAP probe characteristics in Figure~\ref{fig:sweeps}(b)--(d), and in the dashed lines the corresponding $1/r^2$ neutral density profiles. Considering measurement accuracy and the crudeness of the model we do not expect perfect numerical predictions, but a general correspondence is seen. As the cooling rate is finite, we would not expect to see 100\% cold electrons except possibly very close to the nucleus in the most active comet stage, meaning that the point where we find 95\% cold electrons falls inside the boundary of complete cooling is not worrying. Other major limitations of the model include neglect of  electric fields - for example the ambipolar field due to the electron pressure gradient - or fields induced by the solar wind \citep{Madanian2016a,Vigren2015b}, the assumption of fully distinct populations, and the neglect of dissociative recombination, which would operate more efficiently on the cold electron population than on the warm one.

The observation of two electron components at very different temperatures is of interest in itself but also for plasma wave physics at the comet, as this situation allows propagation of electron acoustic waves \citep{Pottelette1999a}.

\subsection{Cold plasma filamentation}\label{sec:filament}

The LAP probe current pulses presented in Section~\ref{sec:cold} were found to be due to localised plasma regions of high density and low electron temperature. Such density variations are not unexpected since they developed in the hybrid simulations for 67P at 1.3~AU by \citet{Koenders2015a}. Figure~6 of that study, which used spherically symmetric outgassing, shows a relatively consistently varying density in the innermost coma, out to some 30-50~km, where the expanding plasma breaks up into filaments or "spikes" of thickness less than 10~km, possibly down to the simulation grid resolution of 2.2~km. The time evolution at a particular point is provided in their Figure~3, where one can find a tenfold increases in ion density over an apparent timescale between a few and a few tens of seconds. \citet{Koenders2015a} noted that these spikes may occur out to 150~km for the particular parameters used in the  simulation. Comparing to our Figure~\ref{fig:odel}, we do find pulses out to at least 400~km at perihelion, but we may note the large increase in pulse numbers as Rosetta comes within about 150~km in mid-November 2015. 

The simulations by \citet{Koenders2015a} are certainly not the first to suggest unstable cometary plasma boundaries. Various authors have discussed the stability of the cometary ionopause and concluded that Kelvin-Helmholtz as well as Rayleigh-Taylor instabilities may set in \citep{Ershkovich1988a,Thomas1995a}. In recent years, MHD simulations by \citet{Rubin2012a}
 in anticipation of Rosetta have indeed shown a Kelvin-Helmholtz instability at a diamagnetic cavity boundary at 2.7~AU, resulting in a filamentation pattern not unlike the one in the hybrid simulations by \citet{Koenders2015a}. Such Kelvin-Helmholtz filamentation may well be the cause of the current pulses we observe with LAP and also of the spikes in the hybrid simulations, though we should note that additional instabilities not present in MHD may be important particularly in the innermost coma, both in reality and in the hybrid simulations. For typical magnetic field strengths of a few tens of nT, water ions at 1~km/s have a gyroradius of a few tens of km, which in the innermost coma (few tens of km) is greater than or comparable to the density scale length even for a perfectly smooth $1/r$ plasma density profile (which sets an upper limit to the possible density scale length equal to $r$). In this situation, other modes, like the lower hybrid drift modes observed in the Earth's magnetotail by \citet{Norgren2012a}, may become the fastest growing modes. \citet{Koenders2015a} noted asymmetries in the filamentation seen in the hybrid simulations { related} to ion gyroradius effects and highlighted the importance of kinetic effects.

\subsection{Pulses and dust
}\label{sec:dust}

Although the ratio of dust to gas mass-loss rate is high for 67P, with a value of around 5 reported by \citet{Rotundi2015a} and \citet{Moreno2016a}, there is as yet little evidence for any strong impact of dust on the plasma. This is in contrasts to what Cassini found in dusty environments in the Saturn E-ring and the Enceladus plume, where a sometimes overwhelming majority of the plasma electrons are attached to dust grains as reported by \citet{Wahlund2009a} and \citet{Morooka2011a} based on the observed difference between ion and electron densities, and later corroborated by good agreement with dust impact measurements on the radio and plasma wave antennas \citep{Engelhardt2015a}. This difference between the environments is not unexpected, as modelling \citep{Vigren2015a} shows that grains bigger than about 0.1~$\mu$m are unlikely to have a significant effect on the overall charge balance in the inner coma. While there are reports of occasional nanograin observations at 67P \citep{Burch2015b,Gombosi2015a} there is as yet no indication of their prevalence, and the evidence so far available is generally in favour of the dust distribution being dominated by larger grains \citep{Rotundi2015a,Fulle2015a,Moreno2016a}. However, one should note that so far there has been no attempt to compare ion and electron densities at 67P at the precision required for inferring the charge density carried by dust, and it is possible that such or other investigations may modify the picture.

It may seem tempting to attribute the LAP current pulses discussed in Section~\ref{sec:cold} to dust grains, and various interpretations of the pulses in terms of dust have also been attempted. We consider first the idea that each pulse is due to a large charged dust grain hitting the probe, slowly depositing its charge over a few seconds. This is not in good agreement with simultaneous observations on LAP1 and LAP2, and also is hard to reconcile with the large amount of charge carried. Dust grains should not charge to much higher potentials $U$ than the spacecraft, typically at $V\ind{s} \approx -10$~V, and the charge $q = C U$ on them may then be estimated from the expression for the vacuum capacitance for a sphere of radius $a$, $C = 4 \pi \epsilon_0 a$, yielding 1~pF per cm radius. For a body at $-10$~V to carry a charge of order $\mu$C to the probe, it would clearly need to be a boulder rather than a grain, even if we allow a capacitance 100 times above vacuum values for a sphere. 

Another possibility is a model where each pulse is due to the arrival of a cloud of much smaller dust grains. Such showers of dust grains have indeed been observed by the GIADA dust analyser on Rosetta \citep{Fulle2015a}. To explain the LAP current pulses with opposite signs of current on the two probes when at opposite potential, we would have to assume the simultaneous presence of grains of positive and negative charge state in a dust cloud. The varying properties of dust grains may in certain environments lead to opposite charging of, for example,\ large and small grains \citep{Horanyi1990a}, though it seems unlikely this would apply over such large regions and long times as we observe pulses. Furthermore, the numbers are hard to fit. If each grain making up a pulse was 10~$\mu$m, we would need 1~billion grains to hit the probe within the $\sim 10$~s lifetime of a pulse to transfer 10~$\mu$C, if assuming vacuum capacitance. The number needed scales inversely with the grain size, but no matter how one plays with the numbers, an explanation of LAP current pulses of the type in Figure~\ref{fig:zoom} as due to charge delivery to the probes by charged dust grains hitting it hardly seems possible.

Another dust-related hypothesis for the LAP probe current pulses would be dust grains with high volatile content settling on spacecraft surfaces and creating a gas cloud by sublimation due to efficient heat transfer from the spacecraft. While this may and should happen at times, a problem here is the long ionisation times expected, leading to low degrees of ionisation in such a gas cloud and correspondingly low signal in the LAP data. The expected photoionisation time is on the order of at least $10^5$~s \citep{Vigren2015b}, so the degree of ionisation in this cloud (whose source would be a few meters away from the probe) should be much lower than in the gas flow from the nucleus (at a typical distance of 10--100~km). Assuming a thermal expansion speed of order 100~m/s for the gas released in the grain sublimation, the fraction of ionisation should be $\lesssim 10^{-7}$. To explain the large charge of $10^{14} e$ over a pulse width $\tau = 10$~s in a current pulse noted above, the number of molecules released would thus need to be at least $N = 10^{21}$ even in the very unlikely case that the probe collects every electron in the cloud. A grain containing 30~mg of water ice could provide this, but the number density $n$ of H$_2$O molecules in the cloud would be enormous; assuming hemispherical expansion at $u = 100$~m/s, we get a H$_2$O number density at $r = 1$~m distance from the source surface of order $n = N/(2 \pi r^2 u \tau) \sim 10^{11}$~cm$^{-3}$, at least two orders of magnitude above the cometary neutral gas density typically observed by ROSINA COPS \citep{Bieler2015a}. While there are some small-scale pulses found in COPS (K.~Altwegg, personal communication), a preliminary investigation has shown very few examples coinciding with LAP signatures, and as yet none with clear pulses of the type in Figure~\ref{fig:zoom}. We thus find no workable dust hypothesis for the majority of the LAP pulses, leaving us with plasma structures as the source of the pulses discussed in Section~\ref{sec:cold}.


\section{Conclusions}\label{sec:concl}

We have in this paper discussed some aspects of the measurements by the RPC-LAP Langmuir probe instrument on Rosetta at comet 67P. Examples of probe characteristics are shown, as well as high-time-resolution measurements of LAP probe current. The data examples were found to be consistent with two electron populations when the neutral density is sufficient for efficient cooling of electrons inside the position of Rosetta. The warm (around 10~eV) population is found throughout the mission and interpreted as electrons retaining the energy acquired at ionisation. The cool fraction needed to fit the few bias sweep examples was found to be consistent with expectations from a simple extension of the Haser model to two electron populations. The cold electron population has not been observed on its own; only together with the warm population. During large parts of the mission, the most conspicuous signature of cold plasma was found to be pulses of high current to the Langmuir probes sampling ions as well as electrons, interpreted as filaments of high density cold plasma released from an inner collisionally dominated plasma region. Alternative explanations of the pulses as due to charged dust were not successful. Electron cooling to temperatures of 0.1~eV or less and filamentation of cometary plasma have both been predicted but not directly observed before Rosetta.


\begin{acknowledgements}
The results presented here are only possible thanks to the combined efforts over 20 years by many groups and individuals involved in Rosetta, including but not restricted to the ESA project teams at ESTEC, ESOC and ESAC and all people involved in designing, building, testing and operating RPC and LAP. We thank Kathrin Altwegg for discussions of the pulses in LAP and COPS. Rosetta is a European Space Agency (ESA) mission with contributions from its member states and the National Aeronautics and Space Administration (NASA). The work on RPC-LAP data was funded by the Swedish National Space Board under contracts 109/12, 171/12, 135/13, 166/14 and 168/15, and by Vetenskapsr{\aa}det under contract 621-2013-4191. This work has made use of the AMDA and RPC Quicklook database, provided by a collaboration between the Centre de Donn\'{e}es de la Physique des Plasmas (CDPP) (supported by CNRS, CNES, Observatoire de Paris and Universit\'{e} Paul Sabatier, Toulouse), and Imperial College London (supported by the UK Science and Technology Facilities Council).
\end{acknowledgements}

%
\bibliographystyle{aa} 
\bibliography{refs} 

\begin{thebibliography}{80}
\expandafter\ifx\csname natexlab\endcsname\relax\def\natexlab#1{#1}\fi

\bibitem[{Balsiger {et~al.}(2007)Balsiger, Altwegg, Bochsler, Eberhardt,
  Fischer, Graf, J\"{a}ckel, Kopp, Langer, Mildner, M¸ller, Riesen, Rubin,
  Scherer, Wurz, W\"{u}thrich, Arijs, Delanoye, Keyser, Neefs, Nevejans,
  R\'{e}me, Aoustin, Mazelle, M\'{e}dale, Sauvaud, Berthelier, Bertaux, Duvet,
  Illiano, Fuselier, Ghielmetti, Magoncelli, Shelley, Korth, Heerlein, Lauche,
  Livi, Loose, Mall, Wilken, Gliem, Fiethe, Gombosi, Block, Carignan, Fisk,
  Waite, Young, \& Wollnik}]{Balsiger2007a}
Balsiger, H., Altwegg, K., Bochsler, P., {et~al.} 2007, Space Sci. Rev., 128,
  745

\bibitem[{Bieler {et~al.}(2015)Bieler, {Fougere, N.}, {Toth, G.}, {Tenishev,
  V.}, {Combi, M.}, {Gombosi, T.}, {Hansen, K.}, {Huang, Z.}, {Jia, X.},
  {Altwegg, K.}, {Wurz, P.}, {Balsiger, H.}, {Jackel, A.}, {Le Roy, L.}, {Gasc,
  S.}, {Calmonte, U.}, {Rubin, M.}, {Tzou, C.-Y.}, {Hassig, M.}, {Fuselier,
  S.}, \& {al, et}}]{Bieler2015a}
Bieler, A., {Fougere, N.}, {Toth, G.}, {et~al.} 2015, A\&A, 583, A7

\bibitem[{Brace(1998)}]{Brace1998a}
Brace, L.~H. 1998, in Measurement Techniques in Space Plasmas: Particles (AGU
  Geophysical Monograph 102), ed. J.~Borovsky, R.~Pfaff, \& D.~Young (American
  Geophysical Union), 23--35

\bibitem[{Broiles {et~al.}(2015)Broiles, {Burch, J.L.}, {Clark, G.B.},
  {Koenders, C.}, {Behar, E.}, {Goldstein, R.}, {Fuselier, S.A.}, {Mandt,
  K.E.}, {Mokashi, P.}, \& {Samara, M.}}]{Broiles2015a}
Broiles, T., {Burch, J.L.}, {Clark, G.B.}, {et~al.} 2015, A\&A, 583, A21

\bibitem[{{Broiles} {et~al.}(2016{\natexlab{a}}){Broiles}, {Burch}, {Chae},
  {Clark}, {Cravens}, {Eriksson}, {Fuselier}, {Frahm}, {Gasc}, {Goldstein},
  {Henri}, {Koenders}, {Livadiotis}, {Mandt}, {Mokashi}, {Nemeth}, {Odelstad},
  {Rubin}, \& {Samara}}]{Broiles2016b}
{Broiles}, T.~W., {Burch}, J.~L., {Chae}, K., {et~al.} 2016{\natexlab{a}},
  MNRAS, 462, S312

\bibitem[{{Broiles} {et~al.}(2016{\natexlab{b}}){Broiles}, {Livadiotis},
  {Burch}, {Chae}, {Clark}, {Cravens}, {Davidson}, {Eriksson}, {Frahm},
  {Fuselier}, {Goldstein}, {Goldstein}, {Henri}, {Madanian}, {Mandt},
  {Mokashi}, {Pollock}, {Rahmati}, {Samara}, \& {Schwartz}}]{Broiles2016a}
{Broiles}, T.~W., {Livadiotis}, G., {Burch}, J.~L., {et~al.}
  2016{\natexlab{b}}, J. Geophys. Res., 121, 7407

\bibitem[{Burch {et~al.}(2015)Burch, Gombosi, Clark, Mokashi, \&
  Goldstein}]{Burch2015b}
Burch, J.~L., Gombosi, T.~I., Clark, G., Mokashi, P., \& Goldstein, R. 2015,
  Geophys. Res. Lett, 6575, 2015GL065177

\bibitem[{Carr {et~al.}(2007)Carr, Cupido, Lee, Balogh, Beek, Burch, Dunford,
  Eriksson, Gill, Glassmeier, Goldstein, Lagoutte, Lundin, Lundin, Lybekk,
  Michau, Musmann, Nilsson, Pollock, Richter, \& Trotignon}]{Carr2007a}
Carr, C., Cupido, E., Lee, C. G.~Y., {et~al.} 2007, Space Sci. Rev., 128, 629

\bibitem[{Coates \& Jones(2009)}]{Coates2009a}
Coates, A. \& Jones, G. 2009, Planet. Space Sci., 57, 1175

\bibitem[{Coates(1997)}]{Coates1997a}
Coates, A.~J. 1997, Adv. Space Res., 20, 255

\bibitem[{{Combi} {et~al.}(2012){Combi}, {Tenishev}, {Rubin}, {Fougere}, \&
  {Gombosi}}]{Combi2012a}
{Combi}, M.~R., {Tenishev}, V.~M., {Rubin}, M., {Fougere}, N., \& {Gombosi},
  T.~I. 2012, ApJ, 749, 29

\bibitem[{{Cravens} {et~al.}(1987){Cravens}, {Kozyra}, {Nagy}, {Gombosi}, \&
  {Kurtz}}]{Cravens1987b}
{Cravens}, T.~E., {Kozyra}, J.~U., {Nagy}, A.~F., {Gombosi}, T.~I., \& {Kurtz},
  M. 1987, J. Geophys. Res., 92, 7341

\bibitem[{{de Almeida} {et~al.}(2009){de Almeida}, Sanzovo, Sanzovo, Boczko, \&
  Torres}]{Almeida2009a}
{de Almeida}, A., Sanzovo, D.~T., Sanzovo, G., Boczko, R., \& Torres, R.~M.
  2009, Adv. Space Res., 43, 1993

\bibitem[{Edberg {et~al.}(2009)Edberg, Eriksson, Auster, Barabash,
  B\"osswetter, Carr, Cowley, Cupido, Fr\"nz, Glassmeier, Goldstein, Lester,
  Lundin, Modolo, Nilsson, Richter, Samara, \& Trotignon}]{Edberg2009a}
Edberg, N., Eriksson, A., Auster, U., {et~al.} 2009, Planet. Space Sci., 57,
  1085

\bibitem[{Edberg {et~al.}(2016{\natexlab{a}})Edberg, Alho, Andr\'{e}, Andrews,
  Behar, Burch, Carr, Cupido, Engelhardt, Eriksson, Glassmeier, Goetz,
  Goldstein, Henri, Johansson, Koenders, Mandt, M\"{o}stl, Nilsson, Odelstad,
  Richter, Simon~Wedlund, Stenberg~Wieser, Szeg\"{o}, Vigren, \&
  Volwerk}]{Edberg2016b}
Edberg, N. J.~T., Alho, M., Andr\'{e}, M., {et~al.} 2016{\natexlab{a}}, MNRAS,
  462, S45

\bibitem[{Edberg {et~al.}(2015)Edberg, Eriksson, Odelstad, Henri, Lebreton,
  Gasc, Rubin, Andre, Gill, Johansson, Johansson, Vigren, Wahlund, Carr,
  Cupido, Glassmeier, Goldstein, Koenders, Mandt, Nemeth, Nilsson, Richter,
  Wieser, Szego, \& Volwerk}]{Edberg2015a}
Edberg, N. J.~T., Eriksson, A.~I., Odelstad, E., {et~al.} 2015, Geophys. Res.
  Lett., 42, 4263

\bibitem[{Edberg {et~al.}(2016{\natexlab{b}})Edberg, Eriksson, Odelstad,
  Vigren, Andrews, Johansson, Burch, Carr, Cupido, Glassmeier, Goldstein,
  Halekas, Henri, Koenders, Mandt, Mokashi, Nemeth, Nilsson, Ramstad, Richter,
  \& Wieser}]{Edberg2016a}
Edberg, N. J.~T., Eriksson, A.~I., Odelstad, E., {et~al.} 2016{\natexlab{b}},
  J. Geophys. Res., 121, 949

\bibitem[{Engelhardt {et~al.}(2015)Engelhardt, Wahlund, Andrews, Eriksson, Ye,
  Kurth, Gurnett, Morooka, Farrell, \& Dougherty}]{Engelhardt2015a}
Engelhardt, I., Wahlund, J.-E., Andrews, D., {et~al.} 2015, Planet. Space Sci.,
  117, 453

\bibitem[{Eriksson {et~al.}(2007)Eriksson, Bostr\"{o}m, Gill, {\AA}hl\'{e}n,
  Jansson, Wahlund, Andr\'{e}, M\"{a}lkki, Holtet, Lybekk, Pedersen, Blomberg,
  \& {the LAP team}}]{Eriksson2007d}
Eriksson, A.~I., Bostr\"{o}m, R., Gill, R., {et~al.} 2007, Space Sci. Rev.,
  128, 729

\bibitem[{Eriksson {et~al.}(2008)Eriksson, Gill, Wahlund, Andr\'{e},
  M\"{a}lkki, Lybekk, Pedersen, Holtet, Blomberg, \& Edberg}]{Eriksson2008a}
Eriksson, A.~I., Gill, R., Wahlund, J.-E., {et~al.} 2008, in Rosetta - ESA's
  Mission to the Origin of the Solar System, ed. R.~Schulz, C.~Alexander,
  H.~Boehnhardt, \& K.-H. Glassmeier (Springer), 435--447

\bibitem[{{Ershkovich} \& {Flammer}(1988)}]{Ershkovich1988a}
{Ershkovich}, A.~I. \& {Flammer}, K.~R. 1988, ApJ, 328, 967

\bibitem[{{Fulle} {et~al.}(2015){Fulle}, {Della Corte}, {Rotundi}, {Weissman},
  {Juhasz}, {Szego}, {Sordini}, {Ferrari}, {Ivanovski}, {Lucarelli}, {Accolla},
  {Merouane}, {Zakharov}, {Mazzotta Epifani}, {L{\'o}pez-Moreno},
  {Rodr{\'{\i}}guez}, {Colangeli}, {Palumbo}, {Gr{\"u}n}, {Hilchenbach},
  {Bussoletti}, {Esposito}, {Green}, {Lamy}, {McDonnell}, {Mennella}, {Molina},
  {Morales}, {Moreno}, {Ortiz}, {Palomba}, {Rodrigo}, {Zarnecki}, {Cosi},
  {Giovane}, {Gustafson}, {Herranz}, {Jer{\'o}nimo}, {Leese},
  {L{\'o}pez-Jim{\'e}nez}, \& {Altobelli}}]{Fulle2015a}
{Fulle}, M., {Della Corte}, V., {Rotundi}, A., {et~al.} 2015, ApJL, 802, L12

\bibitem[{Galand {et~al.}(2016)Galand, H\'{e}ritier, Odelstad, Henri, Altwegg,
  Beth, Broiles, Burch, Carr, Cupido, Eriksson, Glassmeier, Johansson,
  Lebreton, Mandt, Nilsson, Richter, Rubin, Sagni\`{e}res, Schwartz, S\'{e}mon,
  Tzou, Valli\`{e}res, Vigren, \& Wurz}]{Galand2016a}
Galand, M., H\'{e}ritier, K.~L., Odelstad, E., {et~al.} 2016, MNRAS, 462, S331

\bibitem[{Gan \& Cravens(1990)}]{Gan1990a}
Gan, L. \& Cravens, T.~E. 1990, J. Geophys. Res., 95, 6285

\bibitem[{{Garnier} {et~al.}(2012){Garnier}, {Wahlund}, {Holmberg}, {Morooka},
  {Grimald}, {Eriksson}, {Schippers}, {Gurnett}, {Krimigis}, {Krupp}, {Coates},
  {Crary}, \& {Gustafsson}}]{Garnier2012a}
{Garnier}, P., {Wahlund}, J.-E., {Holmberg}, M.~K.~G., {et~al.} 2012, J.
  Geophys. Res., 117, 10202

\bibitem[{{Goetz} {et~al.}(2016){Goetz}, {Koenders}, {Hansen}, {Burch}, {Carr},
  {Eriksson}, {Fr{\"u}hauff}, {G{\"u}ttler}, {Henri}, {Nilsson}, {Richter},
  {Rubin}, {Sierks}, {Tsurutani}, {Volwerk}, \& {Glassmeier}}]{Goetz2016b}
{Goetz}, C., {Koenders}, C., {Hansen}, K.~C., {et~al.} 2016, MNRAS, 462, S459

\bibitem[{Goetz {et~al.}(2016)Goetz, Koenders, Richter, Altwegg, Burch, Carr,
  Cupido, Eriksson, G\"{u}ttler, Henri, Mokashi, Nemeth, Nilsson, Rubin,
  Sierks, Tsurutani, Vallat, Volwerk, \& Glassmeier}]{Goetz2016a}
Goetz, C., Koenders, C., Richter, I., {et~al.} 2016, A\&A, 588, A24

\bibitem[{Goldstein {et~al.}(2015)Goldstein, Burch, Mokashi, Broiles, Mandt,
  Hanley, Cravens, Rahmati, Samara, Clark, Hassig, \& Webster}]{Goldstein2015a}
Goldstein, R., Burch, J.~L., Mokashi, P., {et~al.} 2015, Geophys. Res. Lett.,
  42, 3093

\bibitem[{Gombosi {et~al.}(2015)Gombosi, {Burch, J.L.}, \& {Horanyi,
  M.}}]{Gombosi2015a}
Gombosi, T., {Burch, J.L.}, \& {Horanyi, M.} 2015, A\&A

\bibitem[{Grard {et~al.}(1989)Grard, Laakso, Pedersen, Trotignon, \&
  Michailov}]{Grard1989a}
Grard, R., Laakso, H., Pedersen, A., Trotignon, J.-G., \& Michailov, Y. 1989,
  Ann. Geophysicae, 7, 141

\bibitem[{Grard(1973)}]{Grard1973a}
Grard, R. J.~L. 1973, J. Geophys. Res., 78, 2885

\bibitem[{{Gulkis} {et~al.}(2015){Gulkis}, {Allen}, {von Allmen}, {Beaudin},
  {Biver}, Bockelee-Morvan, {Choukroun}, {Crovisier}, {Davidsson}, {Encrenaz},
  {Encrenaz}, {Frerking}, {Hartogh}, {Hofstadter}, {Ip}, {Janssen}, {Jarchow},
  {Keihm}, {Lee}, {Lellouch}, {Leyrat}, {Rezac}, {Schloerb}, \&
  {Spilker}}]{Gulkis2015a}
{Gulkis}, S., {Allen}, M., {von Allmen}, P., {et~al.} 2015, Science, 347, 709

\bibitem[{{H{\"a}berli} {et~al.}(1996){H{\"a}berli}, {Altwegg}, {Balsiger}, \&
  {Geiss}}]{Haberli1996a}
{H{\"a}berli}, R.~M., {Altwegg}, K., {Balsiger}, H., \& {Geiss}, J. 1996, J.
  Geophys. Res., 101, 15579

\bibitem[{{Hansen} {et~al.}(2016){Hansen}, {Altwegg}, {Berthelier}, {Bieler},
  {Biver}, {Bockel{\'e}e-Morvan}, {Calmonte}, {Capaccioni}, {Combi}, {de
  Keyser}, {Fiethe}, {Fougere}, {Fuselier}, {Gasc}, {Gombosi}, {Huang}, {Le
  Roy}, {Lee}, {Nilsson}, {Rubin}, {Shou}, {Snodgrass}, {Tenishev}, {Toth},
  {Tzou}, {Wedlund}, \& {Rosina Team}}]{Hansen2016a}
{Hansen}, K.~C., {Altwegg}, K., {Berthelier}, J.-J., {et~al.} 2016, MNRAS, 462,
  S491

\bibitem[{{Haser}(1957)}]{Haser1957a}
{Haser}, L. 1957, Bulletin de la Societe Royale des Sciences de Liege, 43, 740

\bibitem[{{H{\"a}ssig} {et~al.}(2015){H{\"a}ssig}, {Altwegg}, {Balsiger},
  {Bar-Nun}, {Berthelier}, {Bieler}, {Bochsler}, {Briois}, {Calmonte}, {Combi},
  {De Keyser}, {Eberhardt}, {Fiethe}, {Fuselier}, {Galand}, {Gasc}, {Gombosi},
  {Hansen}, {J{\"a}ckel}, {Keller}, {Kopp}, {Korth}, {K{\"u}hrt}, {Le Roy},
  {Mall}, {Marty}, {Mousis}, {Neefs}, {Owen}, {R{\`e}me}, {Rubin}, {S{\'e}mon},
  {Tornow}, {Tzou}, {Waite}, \& {Wurz}}]{Hassig2015a}
{H{\"a}ssig}, M., {Altwegg}, K., {Balsiger}, H., {et~al.} 2015, Science, 347,
  276

\bibitem[{Hilgers {et~al.}(1992)Hilgers, Holback, Holmgren, \&
  Bostr\"{o}m}]{Hilgers1992a}
Hilgers, A., Holback, B., Holmgren, G., \& Bostr\"{o}m, R. 1992, J. Geophys.
  Res., 97, 8631

\bibitem[{Holmberg {et~al.}(2012)Holmberg, Wahlund, Morooka, \&
  Persoon}]{Holmberg2012a}
Holmberg, M., Wahlund, J.-E., Morooka, M., \& Persoon, A. 2012, Planet. Space
  Sci., 73, 151

\bibitem[{{Horanyi} \& {Goertz}(1990)}]{Horanyi1990a}
{Horanyi}, M. \& {Goertz}, C.~K. 1990, ApJ, 361, 155

\bibitem[{{Ip} {et~al.}(1986){Ip}, {Schwenn}, {Rosenbauer}, {Balsiger},
  {Neugebauer}, \& {Shelley}}]{Ip1986a}
{Ip}, W.-H., {Schwenn}, R., {Rosenbauer}, H., {et~al.} 1986, in ESA Special
  Publication, Vol. 250, ESLAB Symposium on the Exploration of Halley's Comet,
  ed. B.~{Battrick}, E.~J. {Rolfe}, \& R.~{Reinhard}

\bibitem[{{Itikawa} \& {Mason}(2005)}]{Itikawa2005a}
{Itikawa}, Y. \& {Mason}, N. 2005, Journal of Physical and Chemical Reference
  Data, 34, 1

\bibitem[{Johansson {et~al.}(2016)Johansson, Henri, Eriksson, Vallieres,
  Lebreton, Beghin, Wattieux, \& Odelstad}]{Johansson2016a}
Johansson, F.~L., Henri, P., Eriksson, A., {et~al.} 2016, in Proceedings of the
  14th Spacecraft Charging Technology Conference (European Space Agency)

\bibitem[{Jorda {et~al.}(2016)Jorda, Gaskell, Capanna, Hviid, Lamy, Durech,
  Faury, Groussin, Gutierrez, Jackman, Keihm, Keller, Knollenberg, Kuehrt,
  Marchi, Mottola, Palmer, Schloerb, Sierks, Vincent, {A'Hearn}, Barbieri,
  Rodrigo, Koschny, Rickman, Barucci, Bertaux, Bertini, Cremonese, Deppo,
  Davidsson, Debei, Cecco, Fornasier, Fulle, Guettler, Ip, Kramm, Kueppers,
  Lara, Lazzarin, Moreno, Marzari, Naletto, Oklay, Thomas, Tubiana, \&
  Wenzel}]{Jorda2016a}
Jorda, L., Gaskell, R., Capanna, C., {et~al.} 2016, Icarus, 277, 257

\bibitem[{Karlsson {et~al.}(2017)Karlsson, Eriksson, Odelstad, Nilsson, Kullen,
  Lindqvist, Dickeli, Glassmeier, \& Richter}]{Karlsson2017a}
Karlsson, T., Eriksson, A., Odelstad, E., {et~al.} 2017, Geophys. Res. Lett.,
  44

\bibitem[{{Koenders} {et~al.}(2015){Koenders}, {Glassmeier}, {Richter},
  {Ranocha}, \& {Motschmann}}]{Koenders2015a}
{Koenders}, C., {Glassmeier}, K.-H., {Richter}, I., {Ranocha}, H., \&
  {Motschmann}, U. 2015, Planet. Space Sci., 105, 101

\bibitem[{Laframboise(1966)}]{Laframboise1966a}
Laframboise, J.~G. 1966, Theory of spherical and cylindrical {Langmuir} probes
  in a collisionless, {Maxwellian} plasma at rest, Tech. Rep. UTIAS report 100,
  Institute for Aerospace Studies, University of Toronto

\bibitem[{Laframboise \& Godard(1974)}]{Laframboise1974a}
Laframboise, J.~G. \& Godard, R. 1974, Planet. Space Sci., 22, 1145

\bibitem[{Laframboise \& Parker(1973)}]{Laframboise1973a}
Laframboise, J.~G. \& Parker, L.~W. 1973, Phys. Fluids, 16, 629

\bibitem[{Laframboise \& Sonmor(1993)}]{Laframboise1993a}
Laframboise, J.~G. \& Sonmor, L.~J. 1993, J. Geophys. Res., 98, 337

\bibitem[{Madanian {et~al.}(2016)Madanian, Cravens, Rahmati, Goldstein, Burch,
  Eriksson, Edberg, Henri, Mandt, Clark, Rubin, Broiles, \&
  Reedy}]{Madanian2016a}
Madanian, H., Cravens, T.~E., Rahmati, A., {et~al.} 2016, J. Geophys. Res.,
  121, 2016JA022610

\bibitem[{Mandt {et~al.}(2016)Mandt, Eriksson, Edberg, Koenders, Broiles,
  Fuselier, Henri, Nemeth, Alho, Biver, Beth, Burch, Carr, Chae, Coates,
  Cupido, Galand, Glassmeier, Goetz, Goldstein, Hansen, Haiducek, Kallio,
  Lebreton, Luspay-Kuti, Mokashi, Nilsson, Opitz, Richter, Samara, Szego, Tzou,
  Volwerk, Simon~Wedlund, \& Stenberg~Wieser}]{Mandt2016a}
Mandt, K.~E., Eriksson, A., Edberg, N. J.~T., {et~al.} 2016, MNRAS, 462, S9

\bibitem[{Medicus(1961)}]{Medicus1961a}
Medicus, G. 1961, J. Appl. Phys., 32, 2512

\bibitem[{Meyer-Vernet {et~al.}(1986)Meyer-Vernet, Couturier, Hoang, Perche,
  Steinberg, Fainberg, \& Meetre}]{Meyer-Vernet1986a}
Meyer-Vernet, N., Couturier, P., Hoang, S., {et~al.} 1986, Science, 232, 370

\bibitem[{Moreno {et~al.}(2016)Moreno, {Snodgrass, C.}, {Hainaut, O.},
  {Tubiana, C.}, {Sierks, H.}, {Barbieri, C.}, {Lamy, P. L.}, {Rodrigo, R.},
  {Koschny, D.}, {, et al.}, {Rickman, H.}, {Keller, H. U.}, {Agarwal, J.},
  {A'Hearn, M. F.}, {Barucci, M. A.}, {Bertaux, J.-L.}, {Bertini, I.}, {Besse,
  S.}, {Bodewits, D.}, {Cremonese, G.}, {Da Deppo, V.}, {Davidsson, B.},
  {Debei, S.}, {De Cecco, M.}, {Ferri, F.}, {Fornasier, S.}, {Fulle, M.},
  {Groussin, O.}, {Gutierrez, P. J.}, {Gutierrez-Marques, P.}, {G\"{u}ttler,
  C.}, {Hviid, S. F.}, {Ip, W.-H.}, {Jorda, L.}, {Knollenberg, J.}, {Kovacs,
  G.}, {Kramm, J.-R.}, {K\"{u}hrt, E.}, {K\"{u}ppers, M.}, {Lara, L. M.},
  {Lazzarin, M.}, {López-Moreno, J. J.}, {Marzari, F.}, {Mottola, S.},
  {Naletto, G.}, {Oklay, N.}, {Pajola, M.}, {Thomas, N.}, {Vincent, J. B.},
  {Della Corte, V.}, {Fitzsimmons, A.}, {Faggi, S.}, {Jehin, E.}, {Opitom, C.},
  \& {Tozzi, G.-P.}}]{Moreno2016a}
Moreno, F., {Snodgrass, C.}, {Hainaut, O.}, {et~al.} 2016, A\&A, 587, A155

\bibitem[{Morooka {et~al.}(2011)Morooka, Wahlund, Eriksson, Farrell, Gurnett,
  Kurth, Persoon, Shafiq, Andr\'{e}, \& Holmberg}]{Morooka2011a}
Morooka, M.~W., Wahlund, J.-E., Eriksson, A.~I., {et~al.} 2011, J. Geophys.
  Res., 116, A12221

\bibitem[{Mott-Smith \& Langmuir(1926)}]{Mott-Smith1926a}
Mott-Smith, H.~M. \& Langmuir, I. 1926, Phys. Rev., 28, 727

\bibitem[{{Nilsson} {et~al.}(2015a){Nilsson}, {Stenberg Wieser}, {Behar},
  {Wedlund}, {Gunell}, {Yamauchi}, {Lundin}, {Barabash}, {Wieser}, {Carr},
  {Cupido}, {Burch}, {Fedorov}, {Sauvaud}, {Koskinen}, {Kallio}, {Lebreton},
  {Eriksson}, {Edberg}, {Goldstein}, {Henri}, {Koenders}, {Mokashi}, {Nemeth},
  {Richter}, {Szego}, {Volwerk}, {Vallat}, \& {Rubin}}]{Nilsson2015a}
{Nilsson}, H., {Stenberg Wieser}, G., {Behar}, E., {et~al.} 2015a, Science,
  347, 571

\bibitem[{Nilsson {et~al.}(2015b)Nilsson, {Stenberg Wieser, G.}, {Behar, E.},
  {Simon Wedlund, C.}, {Kallio, E.}, {Gunell, H.}, {Edberg, N. J. T.},
  {Eriksson, A. I.}, {Yamauchi, M.}, {Koenders, C.}, {Wieser, M.}, {Lundin,
  R.}, {Barabash, S.}, {Mandt, K.}, {Burch, J. L.}, {Goldstein, R.}, {Mokashi,
  P.}, {Carr, C.}, {Cupido, E.}, {Fox, P. T.}, {Szego, K.}, {Nemeth, Z.},
  {Fedorov, A.}, {Sauvaud, J.-A.}, {Koskinen, H.}, {Richter, I.}, {Lebreton,
  J.-P.}, {Henri, P.}, {Volwerk, M.}, {Vallat, C.}, \& {Geiger,
  B.}}]{Nilsson2015b}
Nilsson, H., {Stenberg Wieser, G.}, {Behar, E.}, {et~al.} 2015b, A\&A, 583, A20

\bibitem[{Norgren {et~al.}(2012)Norgren, Vaivads, Khotyaintsev, \&
  Andr\'e}]{Norgren2012a}
Norgren, C., Vaivads, A., Khotyaintsev, Y.~V., \& Andr\'e, M. 2012, Phys. Rev.
  Lett., 109, 055001

\bibitem[{Odelstad {et~al.}(2015)Odelstad, Eriksson, Edberg, Johansson, Vigren,
  André, Tzou, Carr, \& Cupido}]{Odelstad2015a}
Odelstad, E., Eriksson, A.~I., Edberg, N. J.~T., {et~al.} 2015, Geophys. Res.
  Lett., 42, 10,126

\bibitem[{Odelstad {et~al.}(2017)Odelstad, Stenberg-Wieser, Wieser, Eriksson,
  Nilsson, \& Johansson}]{Odelstad2017a}
Odelstad, E., Stenberg-Wieser, G., Wieser, M., {et~al.} 2017, MNRAS, in review

\bibitem[{{Odelstad et al.}(2016)}]{Odelstad2016a}
{Odelstad et al.}, E. 2016, in Proceedings of the 14th Spacecraft Charging
  Technology Conference (European Space Agency)

\bibitem[{{Olson} {et~al.}(2010){Olson}, {Brenning}, {Wahlund}, \&
  {Gunell}}]{Olson2010a}
{Olson}, J., {Brenning}, N., {Wahlund}, J., \& {Gunell}, H. 2010, Rev. Sci.
  Instr., 81, 105106

\bibitem[{Pedersen(1995)}]{Pedersen1995a}
Pedersen, A. 1995, Ann. Geophysicae, 13, 118

\bibitem[{{Pottelette} {et~al.}(1999){Pottelette}, {Ergun}, {Treumann},
  {Berthomier}, {Carlson}, {McFadden}, \& {Roth}}]{Pottelette1999a}
{Pottelette}, R., {Ergun}, R.~E., {Treumann}, R.~A., {et~al.} 1999, \grl, 26,
  2629

\bibitem[{{Rotundi} {et~al.}(2015){Rotundi}, {Sierks}, {Della Corte}, {Fulle},
  {Gutierrez}, {Lara}, {Barbieri}, {Lamy}, {Rodrigo}, {Koschny}, {Rickman},
  {Keller}, {L{\'o}pez-Moreno}, {Accolla}, {Agarwal}, {A'Hearn}, {Altobelli},
  {Angrilli}, {Barucci}, {Bertaux}, {Bertini}, {Bodewits}, {Bussoletti},
  {Colangeli}, {Cosi}, {Cremonese}, {Crifo}, {Da Deppo}, {Davidsson}, {Debei},
  {De Cecco}, {Esposito}, {Ferrari}, {Fornasier}, {Giovane}, {Gustafson},
  {Green}, {Groussin}, {Gr{\"u}n}, {G{\"u}ttler}, {Herranz}, {Hviid}, {Ip},
  {Ivanovski}, {Jer{\'o}nimo}, {Jorda}, {Knollenberg}, {Kramm}, {K{\"u}hrt},
  {K{\"u}ppers}, {Lazzarin}, {Leese}, {L{\'o}pez-Jim{\'e}nez}, {Lucarelli},
  {Lowry}, {Marzari}, {Epifani}, {McDonnell}, {Mennella}, {Michalik}, {Molina},
  {Morales}, {Moreno}, {Mottola}, {Naletto}, {Oklay}, {Ortiz}, {Palomba},
  {Palumbo}, {Perrin}, {Rodr{\'{\i}}guez}, {Sabau}, {Snodgrass}, {Sordini},
  {Thomas}, {Tubiana}, {Vincent}, {Weissman}, {Wenzel}, {Zakharov}, \&
  {Zarnecki}}]{Rotundi2015a}
{Rotundi}, A., {Sierks}, H., {Della Corte}, V., {et~al.} 2015, Science, 347,
  aaa3905

\bibitem[{Rubin {et~al.}(2012)Rubin, Hansen, Combi, Daldorff, Gombosi, \&
  Tenishev}]{Rubin2012a}
Rubin, M., Hansen, K.~C., Combi, M.~R., {et~al.} 2012, J. Geophys. Res., 117,
  A06227

\bibitem[{Sj\"{o}gren {et~al.}(2012)Sj\"{o}gren, Eriksson, \&
  Cully}]{Sjogren2012a}
Sj\"{o}gren, A., Eriksson, A.~I., \& Cully, C.~M. 2012, IEEE Trans. Plasma
  Sci., 40, 1257

\bibitem[{Snodgrass {et~al.}(2016)Snodgrass, Opitom, de~Val-Borro, Jehin,
  Manfroid, Lister, Marchant, Jones, Fitzsimmons, Steele, Smith, Jermak,
  Granzer, Meech, Rousselot, \& Levasseur-Regourd}]{Snodgrass2016b}
Snodgrass, C., Opitom, C., de~Val-Borro, M., {et~al.} 2016, MNRAS

\bibitem[{{Tenishev} {et~al.}(2008){Tenishev}, {Combi}, \&
  {Davidsson}}]{Tenishev2008a}
{Tenishev}, V., {Combi}, M., \& {Davidsson}, B. 2008, Astrophys. J., 685, 659

\bibitem[{Thomas(1995)}]{Thomas1995a}
Thomas, V.~A. 1995, J. Geophys. Res., 100, 12017

\bibitem[{{Vigren} {et~al.}(2016){Vigren}, {Altwegg}, {Edberg}, {Eriksson},
  {Galand}, {Henri}, {Johansson}, {Odelstad}, {Tzou}, \&
  {Valli{\'e}res}}]{Vigren2016a}
{Vigren}, E., {Altwegg}, K., {Edberg}, N.~J.~T., {et~al.} 2016, AJ, 152, 59

\bibitem[{Vigren \& Eriksson(2017)}]{Vigren2017a}
Vigren, E. \& Eriksson, A.~I. 2017, AJ, 153, 150

\bibitem[{{Vigren} \& {Galand}(2013)}]{Vigren2013a}
{Vigren}, E. \& {Galand}, M. 2013, ApJ, 772, 33

\bibitem[{Vigren {et~al.}(2015)Vigren, Galand, Eriksson, Edberg, Odelstad, \&
  Schwartz}]{Vigren2015b}
Vigren, E., Galand, M., Eriksson, A.~I., {et~al.} 2015, ApJ, 812, 54

\bibitem[{{Vigren} {et~al.}(2015){Vigren}, {Galand}, {Lavvas}, {Eriksson}, \&
  {Wahlund}}]{Vigren2015a}
{Vigren}, E., {Galand}, M., {Lavvas}, P., {Eriksson}, A.~I., \& {Wahlund},
  J.-E. 2015, ApJ, 798, 130

\bibitem[{Wahlund {et~al.}(2009)Wahlund, Andre, Eriksson, Lundberg, Morooka,
  Shafiq, Averkamp, Gurnett, Hospodarsky, Kurth, Jacobsen, Pedersen, Farrell,
  Ratynskaia, \& Piskunov}]{Wahlund2009a}
Wahlund, J.-E., Andre, M., Eriksson, A., {et~al.} 2009, Planet. Space Sci., 57,
  1795

\bibitem[{Wahlund {et~al.}(2005)Wahlund, Bostr\"{o}m, Gustafsson, Gurnett,
  Kurth, Averkamp, Hospodarsky, Persoon, Canu, Pedersen, Desch, Eriksson, Gill,
  Morooka, \& Andr\'{e}}]{Wahlund2005a}
Wahlund, J.-E., Bostr\"{o}m, R., Gustafsson, G., {et~al.} 2005, Geophys. Res.
  Lett., 32, doi:10.1029/2005GL022699

\bibitem[{Wang {et~al.}(2015)Wang, Hsu, \& Horanyi}]{Wang2015a}
Wang, X., Hsu, H.-W., \& Horanyi, M. 2015, J. Geophys. Res., 120, 2428,
  2014JA020624

\bibitem[{Yang {et~al.}(2016)Yang, Paulsson, Simon~Wedlund, Odelstad, Edberg,
  Koenders, Eriksson, \& Miloch}]{Yang2016a}
Yang, L., Paulsson, J. J.~P., Simon~Wedlund, C., {et~al.} 2016, MNRAS, 462, S33

\end{thebibliography}
%

\end{document}